%% file: M31-DM.tex
\RequirePackage[left]{lineno}
\documentclass[aps,prd,superscriptaddress,floatfix,nofootinbib,notitlepage,reprint]{revtex4-1}

\usepackage{graphicx}
\usepackage{amsmath}
\usepackage{longtable}
\usepackage{aas_macros}
\usepackage{xcolor}
\usepackage{multirow}
\usepackage{mathrsfs}

\graphicspath{{./figures/}}

\usepackage[sort&compress]{natbib}
\usepackage{hyperref}
\usepackage{url}
\begin{document}

% Title
\title{Search for Dark Matter Gamma-ray Emission from the Andromeda Galaxy with the High-Altitude Water Cherenkov Observatory}

\date{\today}

% Author list
\input{authors.tex}

\begin{abstract}
The Andromeda Galaxy (M31) is a nearby ($\sim$780 kpc) galaxy similar to our own Milky Way. Observational evidence suggests that it resides in a large halo of dark matter (DM), making it a good target for DM searches. We present a search for gamma rays from M31 using 1017 days of data from the High Altitude Water Cherenkov (HAWC) Observatory. With its wide field of view and constant monitoring, HAWC is well-suited to search for DM in extended targets like M31. No DM annihilation or decay signal was detected for DM masses from 1 to 100 TeV in the $b\bar{b}$, $t\bar{t}$, $\tau^{+}\tau^{-}$, $\mu^{+}\mu^{-}$, and $W^{+}W^{-}$ channels. Therefore we present limits on those processes. Our limits nicely complement the existing body of DM limits from other targets and instruments. Specifically the DM decay limits from our benchmark model are the most constraining for DM masses from 25 TeV to 100 TeV in the $b\bar{b}$ and $t\bar{t}$ channels. In addition to DM-specific limits, we also calculate general gamma-ray flux limits for M31 in 5 energy bins from 1 TeV to 100 TeV.

\end{abstract}
\maketitle

\input{intro.tex}

\input{dm.tex}

\input{analysis.tex}

\input{results.tex}

\input{conclusions.tex}

\begin{acknowledgments}

We thank the CLUMPY development team for providing us with the developer's version of the software that contained the most recent models.
We also thank Miguel Angel Sanchez Conde for useful discussions regarding the DM halo modeling and Pasquale Serpico for useful discussions on the absorption of high energy gamma rays.
We acknowledge the support from: the US National Science Foundation (NSF); the
US Department of Energy Office of High-Energy Physics; the Laboratory Directed
Research and Development (LDRD) program of Los Alamos National Laboratory;
Consejo Nacional de Ciencia y Tecnolog\'{\i}a (CONACyT), M{\'e}xico (grants
271051, 232656, 260378, 179588, 239762, 254964, 271737, 258865, 243290,
132197), Laboratorio Nacional HAWC de rayos gamma; L'OREAL Fellowship for
Women in Science 2014; Red HAWC, M{\'e}xico; DGAPA-UNAM (grants IG100317,
IN111315, IN111716-3, IA102715, 109916, IA102917); VIEP-BUAP; PIFI 2012, 2013,
PROFOCIE 2014, 2015;the University of Wisconsin Alumni Research Foundation;
the Institute of Geophysics, Planetary Physics, and Signatures at Los Alamos
National Laboratory; Polish Science Centre grant DEC-2014/13/B/ST9/945;
Coordinaci{\'o}n de la Investigaci{\'o}n Cient\'{\i}fica de la Universidad
Michoacana. Thanks to Luciano D\'{\i}az and Eduardo Murrieta for technical support.

\end{acknowledgments}
\clearpage

% Specify following sections are appendices. 
\appendix
\begin{widetext}
\input{app1.tex}
\end{widetext}
\clearpage

\bibliography{M31_DM}

\end{document}

%% file: authors.tex
%\author{Fluttershy}
%\email{kindness@lanl.gov}
%\affiliation{Ponyville}

\author{A.~Albert}
\email{amalbert@lanl.gov}
\address{Physics Division, Los Alamos National Laboratory, Los Alamos, NM, USA }

\author{R.~Alfaro}
\address{Instituto de F\'{i}sica, Universidad Nacional Aut\'{o}noma de M\'{e}xico, Ciudad de Mexico, Mexico }

\author{C.~Alvarez}
\address{Universidad Aut\'{o}noma de Chiapas, Tuxtla Guti\'{e}rrez, Chiapas, M\'{e}xico}

\author{J.D.~\'{A}lvarez}
\address{Universidad Michoacana de San Nicol\'{a}s de Hidalgo, Morelia, Mexico }

\author{R.~Arceo}
\address{Universidad Aut\'{o}noma de Chiapas, Tuxtla Guti\'{e}rrez, Chiapas, M\'{e}xico}

\author{J.C.~Arteaga-Vel\'{a}zquez}
\address{Universidad Michoacana de San Nicol\'{a}s de Hidalgo, Morelia, Mexico }

\author{D.~Avila Rojas}
\address{Instituto de F\'{i}sica, Universidad Nacional Aut\'{o}noma de M\'{e}xico, Ciudad de Mexico, Mexico }

\author{H.A.~Ayala Solares}
\address{Department of Physics, Pennsylvania State University, University Park, PA, USA }

\author{A.~Becerril}
\address{Instituto de F\'{i}sica, Universidad Nacional Aut\'{o}noma de M\'{e}xico, Ciudad de Mexico, Mexico }

\author{E.~Belmont-Moreno}
\address{Instituto de F\'{i}sica, Universidad Nacional Aut\'{o}noma de M\'{e}xico, Ciudad de Mexico, Mexico }

% ...

%% file: intro.tex
\section{INTRODUCTION}\label{sec:intro}

There is ample evidence, from the early Universe to the present, that suggests the majority of matter is composed of a new substance called dark matter (DM). DM is theorized to be a particle that exists outside the Standard Model of particle physics~\citep{Ade:2013zuv,Clowe:2006eq,Sofue:2000jx}. The observational evidence for DM is based primarily on its gravitational influence making the particle nature (e.g. mass, interaction strength) of DM elusive.

Many particle candidates for DM are proposed, such as Weakly Interacting Massive Particles (WIMPs)~\citep{Feng:2010gw,Baer:2014eja}, and are predicted to annihilate or decay to Standard Model particles. These annihilations or decays are expected to produce a pair of Standard Model particles most of whom fragment and produce showers of secondary particles including gamma rays. This results in a continuum in energy of gamma rays in addition to other particles coming from their regions. We can search for these gamma rays with the High Altitude Water Cherenkov (HAWC) Observatory, which observes 2/3 of the sky from $\sim$500 GeV to $\sim$100 TeV every day. Since gamma rays from the local group are not noticeably scattered on their way to Earth, we can use HAWC to search for gamma-ray excesses from known DM targets, several of which lie in the HAWC field of view. While other DM candidates exist (e.g. primodial black holes~\citep{Carr:2016drx,Bird:2016dcv}, axions~\cite{Peccei:1977hh}, and axion-like-particles~\cite{Arias:2012az}), TeV WIMPs and WIMP-like particles are well motivated and worth searching for.

The current best limits for canonical WIMP masses (10-100 GeV) are from the \textit{Fermi} Large Area Telescope (\textit{Fermi} LAT) Collaboration gamma-ray search in 15 dwarf spheriodal galaxies~\citep{Ackermann:2015zua}. These limits exclude WIMPs that were in thermal equilibrium in the early Universe for masses below 100 GeV for particle DM annihilation to a pair of b quarks ($b\bar{b}$) or tau leptons ($\tau^+\tau^-$). This, along with a lack of DM detection at the Large Hadron Collider (e.g.~\citep{Aaboud:2017dor,Aaboud:2016obm,Sirunyan:2017onm,Khachatryan:2016mdm}) or in underground direct detection experiments (e.g.~\citep{Akerib:2017kat,Aprile:2016swn}), motivates searches for TeV gamma rays from higher mass DM annihilation or decay~\cite{Blanco:2017sbc,Garcia-Cely:2015dda,Cholis:2008qq}. 

A good DM target for HAWC is the Andromeda Galaxy (M31) given that it is only $\sim$780 kpc away, has a large inferred dark matter content~\citep{tamm2012stellar}, and that it resides in the HAWC field of view. Additionally, with HAWC's large field of view ($\sim$ 2 sr), we can uniquely observe the majority of the extended M31 DM halo at TeV energies. Similar to the Milky Way DM halo, M31 is known to have several dwarf galaxies~\cite{Tollerud:2011mi}. These clumps of DM (or subhalos) are evidence of the existence of substructure within the M31 DM halo. Therefore one must model both the underlying smooth component and the substructure.
%We define several DM halo models that span the theoretically allowed parameter space for the M31 dark matter halo. Using these spatial models we search signal from DM annihilation or decay into the following channels: $b\bar{b}$, $t\bar{t}$, $\tau^{+}\tau^{-}$, $\mu^{+}\mu^{-}$, and $W^{+}W^{-}$.

M31 is detected at GeV energies by the \textit{Fermi} LAT~\citep{M31Fermi}. This is expected since cosmic ray interactions are predicted to also create gamma rays. However, a recent paper explored the implications of interpreting the \textit{Fermi}-LAT signal as coming from DM~\citep{McDaniel:2018vam}.

%We simply model the emission from M31 as coming from DM only. This is because the non-DM backgrounds are expected to fall steeply at TeV energies. Even extrapolating from the best fit to the \textit{Fermi} LAT data\cite{M31Fermi}, the expected flux at TeV energies is much lower than the HAWC sensitivity. Also, not including an additional non-DM component in our fits makes our derived limits conservative.

In Section~\ref{sec:dm}, we describe the DM models used in this search. 
In Section~\ref{sec:analysis}, we discuss the HAWC instrument, the data set, and the data analysis techniques used.
We show results in~\ref{sec:results}, and conclude in Section~\ref{sec:conclusion}.

%% file: dm.tex
\section{M31 DARK MATTER MODELING}\label{sec:dm}

The expected gamma-ray flux for a given angular region of interest (ROI, $\Delta\Omega$) from DM annihilation ($\frac{d\phi_{Ann}}{dE_\gamma}$, Eq.~\ref{eq:dmfluxA}) and decay ($\frac{d\phi_{Dec}}{dE_\gamma}$, Eq.~\ref{eq:dmfluxD}) is  

\footnotesize\begin{equation}
\frac{d\phi_{Ann}}{dE_\gamma} = \left(\frac{\langle \sigma v\rangle}{8\pi} \frac{dN_\gamma}{dE_\gamma} \frac{1}{m_{DM}^2} \right)\left(\int_{\Delta\Omega} d\Omega	\int_{\rm l.o.s.}d\ell \rho^2_{DM}(\vec{\ell}) \right) \label{eq:dmfluxA}
\end{equation}
\normalsize and
\footnotesize\begin{equation}
\frac{d\phi_{Dec}}{dE_\gamma} = \left(\frac{1}{\tau_{DM}}\frac{1}{4\pi} \frac{dN_\gamma}{dE_\gamma} \frac{1}{m_{DM}} \right)\left(\int_{\Delta\Omega} d\Omega	\int_{\rm l.o.s.}d\ell \rho_{DM}(\vec{\ell}) \right) \label{eq:dmfluxD}
\end{equation}
\normalsize where $\langle \sigma v\rangle$ is the velocity-averaged DM annihilation cross section, $\frac{dN_\gamma}{dE_\gamma}$ is the channel-specific gamma-ray differential spectrum, $m_{DM}$ is the DM mass, $\rho_{DM}$ is the DM density, and $\tau_{DM}$ is the DM lifetime. 

Each equation is composed of a spectral term and a spatial term in the right and left parentheses respectively.  The spectral term for a given DM mass and channel is derived using the DMSpectra class in the Multi-Mission Maximum Likelihood (3ML) software\footnote{\url{https://github.com/giacomov/3ML}}~\citep{Vianello:2015wwa}. It is identical to the \textit{Fermi}-LAT tool DMFitFunction~\citep{Jeltema:2008hf}\footnote{\url{https://fermi.gsfc.nasa.gov/ssc/data/analysis/scitools/source_models.html}} for $m_{DM} < 10$ TeV. For $m_{DM} > 10$ TeV, the annihilation and decay spectra are those from the HAWC dwarf spheroidal study~\citep{Albert:2017vtb}. The gamma-ray energy spectra are derived using Pythia~\citep{Sjostrand:2006za,Sjostrand:2014zea} where both initial particles from DM annihilation or decay have an energy of $m_{DM}$ and $\frac{m_{DM}}{2}$ respectively. This is because we assume that DM is cold and the center of mass energy is simply the DM mass energy. We note that we only consider gamma rays from prompt emission, though additional gamma rays produced during secondary inverse Compton scattering could effect the leptonic channels spectra~\cite{Esmaili:2015xpa}. This makes our results conservative.

The spatial term, called the $J-$factor or $D-$factor for annihilation and decay respectively, is determined by the modeling of the DM halo in M31. It involves an integral over the line of sight (l.o.s) of $\rho_{DM}^2$ for DM annihilation and $\rho_{DM}$ for DM decay as a function of position $\vec{\ell}$. One typically assumes  a spectral model (e.g. 1 TeV mass DM annihilating or decaying to a pair of b quarks) and a DM halo model and solve for either $\langle \sigma v\rangle$ or $\tau_{DM}$ given an observed gamma-ray flux. For a given spectral model, the expected gamma-ray flux from DM annihilation (decay) is proportional to the $J-$factor ($D-$factor). Therefore targets with larger $J-$factors ($D-$factors) probe $\langle \sigma v\rangle$ ($\tau_{DM}$) more deeply.

We simply model the emission from M31 as coming from DM only. This is because the non-DM backgrounds are expected to fall steeply at TeV energies. Even extrapolating from the best fit to the \textit{Fermi}-LAT data~\cite{M31Fermi}, the expected flux at TeV energies is much lower than the HAWC sensitivity. Also, not including an additional non-DM component in our fits makes our results conservative.

To account for the uncertainty of the DM density distribution (and therefore the $J-$ and $D-$factors) of M31, we define two limiting DM templates (MAX and MIN) that represent some of the most optimistic and pessimistic models of DM clustering in an M31 sized galaxy. MIN and MAX produce the smallest and largest $J-$factors respectively. We also create a benchmark template (MED) that is the model best representing recent observations and simulations. 

We use the public code \textsc{\textcolor{black}{CLUMPY}} ~\citep{bonnivard2016clumpy,charbonnier2012clumpy} to generate $J-$factor and $D-$factor maps of M31 for annihilating or decaying DM given the model parameters. From CLUMPY, we generate $14^{\circ}\times 14^{\circ}$ $J-$factor and $D-$factor maps. This corresponds to a radius of $7^{\circ}$ from the halo center, which is where the $J-$factor decreases by approximately 2 orders of magnitude. Below we describe the MIN, MED, and MAX model parameters which are summarized in Tab.~\ref{tab:substructure_para}.

Our M31 DM halo models contain both a smooth component and a substructure component. To define the smooth component of our DM halo models, we use the parameters of the fits of various DM density profiles to the observed M31 stellar velocity curves reported in Ref~\citep{tamm2012stellar}. The DM profiles are spherically symmetric and peak towards the center of M31, but their inner slopes are not well constrained by the stellar velocities.

In addition to the smooth halo, we know that smaller overdensities of DM exist from observations of dwarf galaxies in the M31 DM halo~\cite{Tollerud:2011mi}. Recent high-resolution N-body simulations of spiral galaxies like the Milky Way and M31 also reveal the existence of smaller DM halos (subhalos) within the larger smooth DM halo~\citep{springel2008aquarius,kuhlen2008via,griffen2016caterpillar}. This substructure can lead to a substantial boost of the gamma-ray signal from sources like M31 in case of annihilating DM
~\cite{kuhlen2008dark,Sanchez-Conde:2013yxa} (decaying DM is not very sensitive to subhalos). For the substructure, we need to specify several parameters that govern the amount and distribution of subhalos in the DM halo. The parameters with the largest impact on the expected $J-$factors and $D-$factors are:
\begin{itemize}
\item [-] The index $\alpha$ of the subhalo mass function $\textrm{d}N/\textrm{d}M$ 
\item [-] the fraction of the DM halo mass which is stored in substructure, $f_{\textrm{sub}}=M_{sub} / M_{halo}$,
\item [-] the minimal mass of DM subhalos $M_{\textrm{min}}$, and
\item [-] the concentration of subhalos $c_{\rm{sub}}$ ~\cite{bullock2001profiles,moline2017characterization}.
\end{itemize}
Recent N-body simulations of DM halos suggest that the subhalo mass function is a power-law ($\textrm{d}N/\textrm{d}M \propto M^{-\alpha}$)~\cite{kuhlen2008via,springel2008aquarius}. Large values of $\alpha$ result in larger $J$-factors. If a larger fraction of the total mass of the halo is in substructure (larger $f_{\textrm{sub}}$), the $J-$factor is larger. This is because subhalos not located in the very center of the halo are more dense than the smooth component and contribute more significantly. Therefore more substructure gives a larger boost to the total predicted gamma-ray flux from DM annihilation since the $J-$factor is proportional to an integral over $\rho_{\rm{DM}}^2$. The resolution of N-body simulations typically constrain $M_{\textrm{min}}$. Smaller values of $M_{\textrm{min}}$ result in more subhalos and therefore larger $J-$factors.

The concentration, $c_{\rm{sub}}$, describes the DM density in subhalos for a given subhalo mass and radial distance from the center. It is determined by the subhalo DM density profile (assumed to be a Navarro-Frenk-White profile~\cite{Navarro:1995iw}) and total subhalo mass. Larger mass halos tend to be less concentrated than smaller mass halos. For $c_{\rm{sub}}$ we rely on the most recent model of the concentration parameter of subhalos~\citep{moline2017characterization}\footnote{To this end, we make use of a developer's version of CLUMPY which already features this concentration model.} . This model reports a flattening of the concentration of subhalos towards the low-mass tail of the relation and, furthermore, it includes a dependence on the position of the subhalo within its host halo. We use this concentration parameter model for all three DM halo models.

Finally, we need to decide if the radial distribution of subhalos inside their host halo follows the smooth DM density profile (biased subhalo distribution) or if we want to account for tidal disruption and other effects in the inner regions that would spoil a biased behaviour. This case is usually called an antibiased subhalo distribution and would decrease the expected gamma-ray emission from M31. However, the authors of Ref~\cite{moline2017characterization} argue that tidal disruption of subhalos does not diminish the boost from substructure to a large extent. Thus, we use a biased subhalo distribution for each of our three DM templates of M31.

\begin{figure}[ht]
    \centering
    \includegraphics[width=0.48\textwidth]{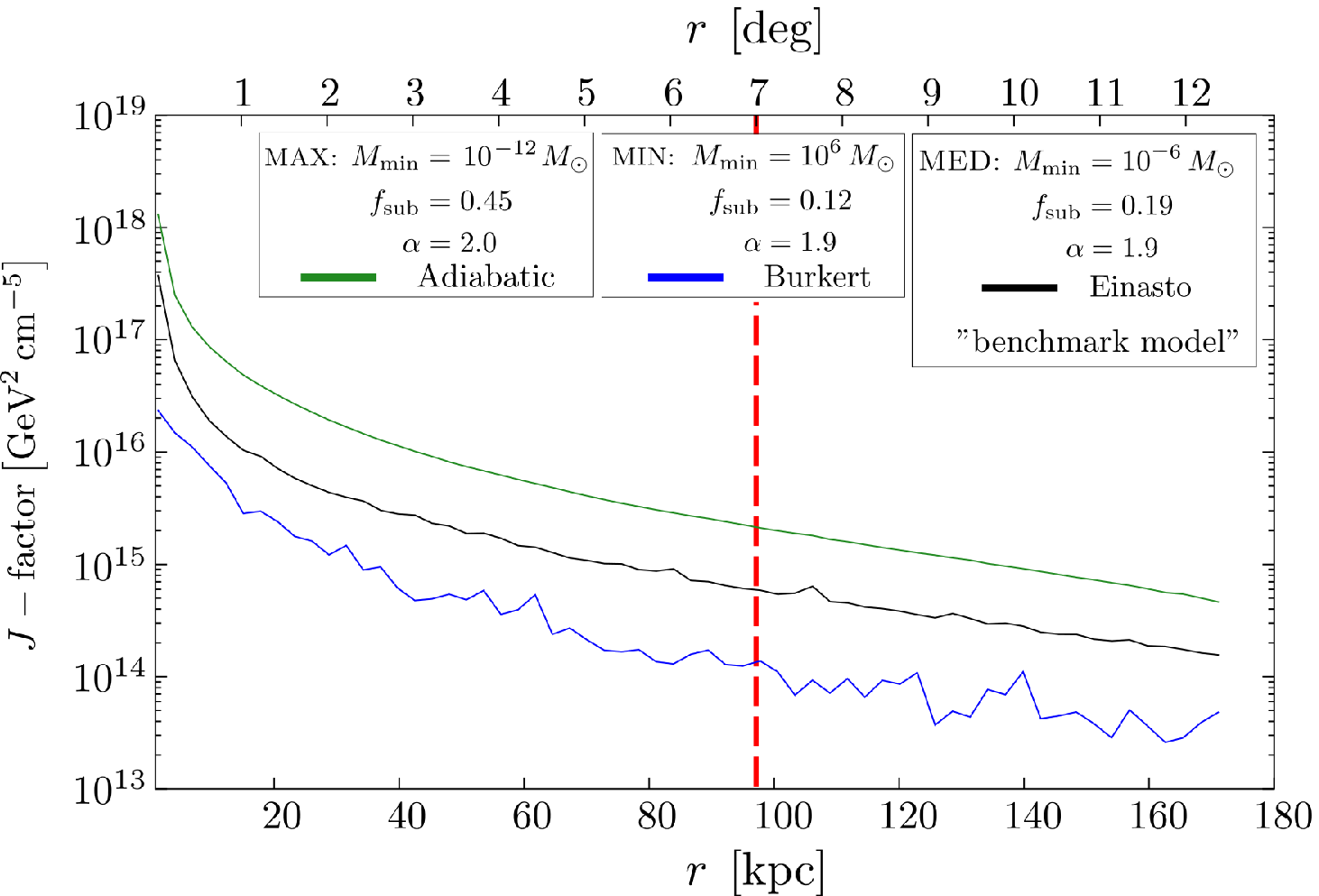}
    \includegraphics[width=0.48\textwidth]{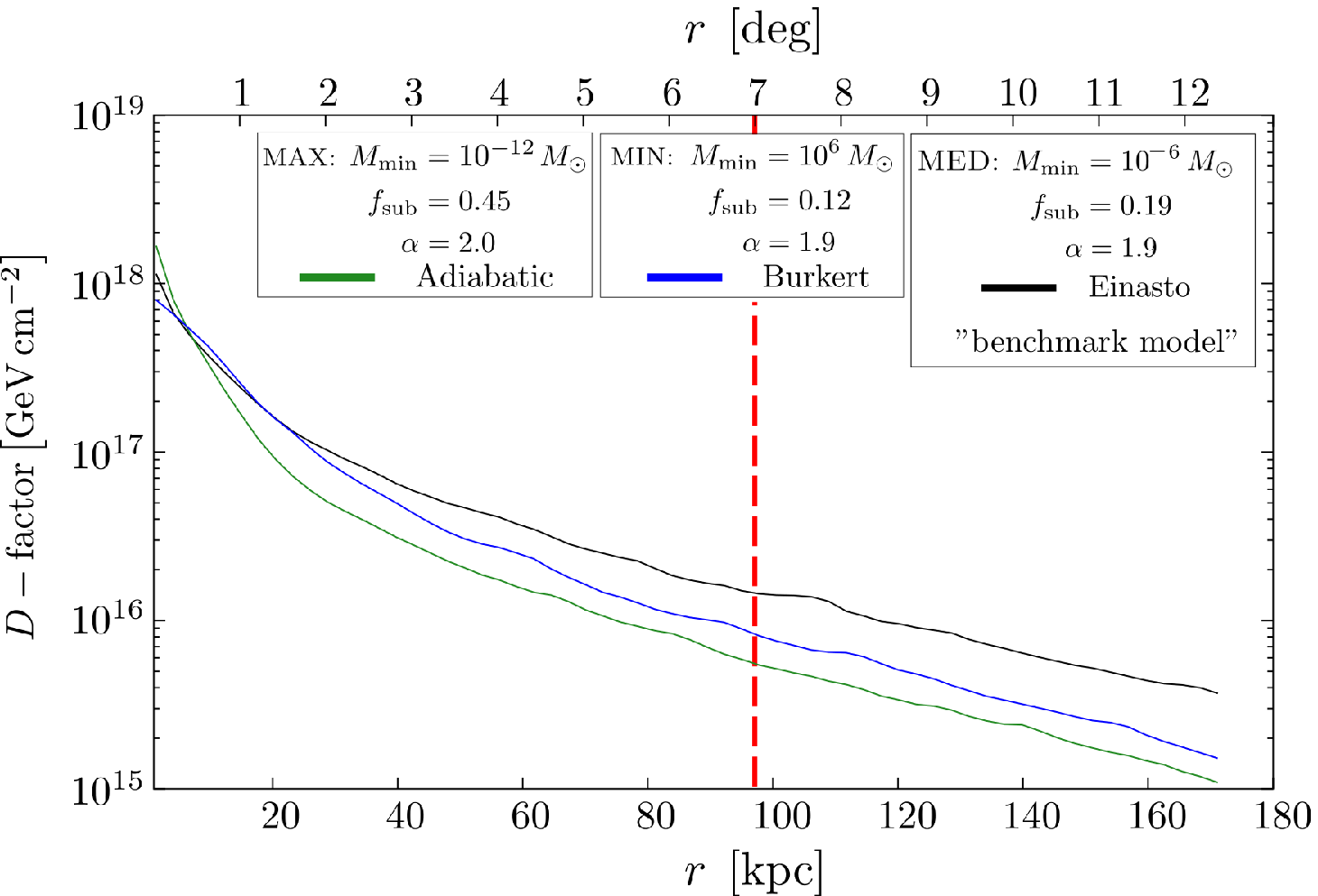} 
    \caption{Radial profiles of the $J-$factors (top)
and $D-$factors (bottom) using the three DM
templates of M31. The red dashed line shows the edge of our region of interest. \label{fig:d-and-j-factors}}
  \end{figure}

We show in Fig.~\ref{fig:d-and-j-factors} the generated
radial profiles of the $J-$ and $D-$factors using the models defined
below and whose main parameters we summarize in Tab.~\ref{tab:substructure_para}. The total $J-$ and $D-$factors in our ROI are given in Tab.~\ref{tab:jfactors}. Note that the MED and MIN D-factors are larger than the MAX D-factor since the virial halo masses (M$_{\rm{vir}}$) are larger.

\subsection{MIN DM Halo Model}

For the smooth halo component we find that the Burkert~\citep{Burkert:1995yz} profile yields the smallest total $J-$factor because it has a nearly constant inner DM density. This is because it is more cored (less cuspy) than others considered in Ref~\citep{tamm2012stellar}. Additionally, the best fit of the Burkert profile to the velocity rotation curves results in a smaller total M31 halo mass.

A recent N-body simulation of DM halos, the Caterpillar simulation (2015) ~\cite{griffen2016caterpillar}, simulated 24 Milky Way sized halos. The authors found that about 12\% of the total halo mass is stored in its substructure ($f_{sub}=0.12$) while they achieve a resolution of $\mathcal{O}\left(10^{4}\,M_{\odot}\right)$ for subhalos. The best fit value of the subhalo mass function's index ($\alpha$) in the Caterpillar simulation is given by $\alpha=1.9\pm0.10$. Therefore we adopt $\alpha=1.9$.

For our MIN model we choose $M_{\textrm{min}}=10^{6}\,M_{\odot}$
as this value is the upper limit for this parameter in CLUMPY and
it is comparable to the mass of the least massive dwarf spheroidal
galaxies~\citep{strigari2007redefining,walker2007velocity}.

\subsection{MED (benchmark) DM Halo Model}

Our MED model uses the parameters of current best fit observations and simulations of the M31 DM halo. For the smooth component, we choose the Einasto profile~\citep{Navarro:2008kc}, which is a cuspy profile that rises towards the galactic center. It is less cuspy than the canonical Navarro-Frenk-White profile~\citep{Navarro:1995iw}. However observations of spiral galaxies suggest that the central cusp is not as steep as the Navarro-Frenk-White profile~\citep{Simon:2004sr,Weldrake:2002ri}. In addition this profile fits well to recent N-body simulations~\citep{2011AJ....142..109C,Navarro:2008kc} and also is not ruled out by M31 rotation curves~\citep{tamm2012stellar}.

For $f_{sub}$ we use a slightly larger value than in the MIN model ($f_{sub}=0.19$) for our MED model. This is the value found in the Aquarius simulation~\cite{springel2008aquarius}. For $\alpha$ we use the best fit value from the Caterpillar simulation ($\alpha=1.9$). We relax the extreme value of $M_{\textrm{min}}=10^{-12}\,M_{\odot}$ used in the MAX case to $M_{\textrm{min}}=10^{-6}\,M_{\odot}$ which is frequently used in the literature.

\subsection{MAX DM Halo Model}

In the MAX model, we model the smooth component as an adiabatically contracted profile. Since this profile rises steeply towards the galactic center, it results in the largest $J-$factor. M31 seems to be the only well-studied galaxy which showed evidence of adiabatic contraction around its central region~\cite{gnedin2011halo}. We adopt the model ``M1 B86'' of~\cite{seigar2008revised}, which is the best-fitting model to the H$\alpha$ rotation curve from Ref~\citep{Rubin:1970zza}. We determine the smooth DM density profile by reading off the ``Halo'' mass-to-radius curve in their Fig.~6 and converting it into a radial density profile via $\rho(r)=\left(4\pi r^{2}\right)^{-1}\textrm{d}M/\textrm{d}r$. 

While for the MIN models we used $f_\textrm{sub}=0.12$, this was based on the Caterpillar simulation with subhalo mass resolution of $\mathcal{O}\left(10^{4}\,M_{\odot}\right)$. However it has been shown
that the minimal subhalo mass depends on the particle physics nature
of a DM particle so that it can cover more orders of magnitude, even values down to $10^{-12}\,M_{\odot}$~\cite{binder2017early,bringmann2009particle}. Previous $N-$body simulations like the Aquarius project or the Via Lactae simulation ~\citep{springel2008aquarius,kuhlen2008via} extrapolated
their results down to smaller subhalo masses and found that at
most 45\% of the total DM halo mass can be present in form of substructure. We therefore use $f_{sub}=0.45$ for our MAX model.

We take the upper range of the best fit value of $\alpha$ from the Caterpillar simulation ($\alpha=1.9\pm0.1$) for our MAX model: $\alpha=2.0$. We also define the smallest subhalo mass to be $10^{-12}\,M_{\odot}$, which results in the largest $J-$factor.

\begin{table*}
\begin{tabular}{ccccc|c}
\hline
\hline
DM Halo Model & smooth profile & $\alpha$ & $f_{\textrm{sub}}$ & $M_{\textrm{min}}\;\left[M_{\odot}\right]$ & M$_{\rm{vir}}\left[10^{10}M_{\odot}\right]$ \tabularnewline
\hline 
MIN & Burkert & $1.9$ & $0.12$ & $10^{6}$ & 79\tabularnewline
MED & Einasto & $1.9$ & $0.19$ & $10^{-6}$ & 113\tabularnewline
MAX & adiabatically contracted NFW & $2.0$ & $0.45$ & $10^{-12}$ & 57\tabularnewline
\end{tabular}
\caption{Summary of the most important parameters of CLUMPY to model the substructure contribution to the total $J-$factor and $D-$factor. Also shown is the halo mass for each model from Refs.~\cite{tamm2012stellar,seigar2008revised}.\label{tab:substructure_para}}
\end{table*}

\begin{table*}[ht]
\begin{tabular}{ccccccc}
\hline \hline
DM Halo Model & J-factor [GeV$^{2}$cm$^{-5}$] & D-factor [GeV cm$^{-2}$] \\
\hline
MIN & $2.00\times10^{19}$ & $1.24\times10^{20}$  \\
MED & $1.27\times10^{20}$ & $1.56\times10^{20}$  \\
MAX & $5.03\times10^{20}$ & $8.95\times10^{19}$  \\
\end{tabular}
\caption{The total $J-$factor and $D-$factor for each DM halo model in a 14$^{\circ}$x14$^{\circ}$ region. Note the $D-$factor mostly depends on the halo mass.
\label{tab:jfactors}}
\end{table*}

%% file: analysis.tex
\section{DETECTOR, DATA, AND ANALYSIS}\label{sec:analysis}

For this analysis we use a 1017 day HAWC dataset from November 26 2014 to December 20 2017. HAWC is a wide field of view survey instrument that scans ~2/3 of the sky each day from $\sim$500 GeV up to $\sim$100 TeV~\citep{Abeysekara:2017mjj}. It consists of 300 large light-tight tanks of water. High energy cosmic particles (e.g. protons and gamma rays) produce showers of secondary particles in the atmosphere that are detected in the tanks via Cherenkov radiation. The full HAWC array was completed in March of 2015. HAWC operates day and night during any weather with a $>$90\% duty cycle. HAWC is located in Sierra Negra, Mexico at an altitude of 4100m at latitude $18^{\circ}59.7' N$ and longitude $97^{\circ}18.6' W$. HAWC observes extensive air showers initiated by high energy particles in the atmosphere. HAWC's angular resolution and background suppression depend on the number of photomultiplier tubes hit, so we bin the data according to what fraction of the available photomultiplier tubes were hit (see Ref.~\citep{Abeysekara:2017mjj} Table 2 for exact bin definitions). We use analysis bins 1-9. These analysis bins correlate with energy, but still have large energy dispersions (see Fig. 3 of~\citep{Abeysekara:2017mjj}). More details on the HAWC detector can be found in Ref.~\citep{Abeysekara:2017mjj}.  

We perform a likelihood ratio test using the 3ML software~\citep{Vianello:2015wwa}. Specifically the likelihood for the signal and null hypotheses is the Poisson distribution in each bin.

\begin{equation}
\mathscr{L} = \Pi_{i,j} \frac{(B_{i,j}+S_{i,j})^{N_{i,j}}\rm{exp}[-B_{i,j}+S_{i,j}]}{N_{i,j}!}
\end{equation}

\noindent where $B_{i,j}$ is the number of background counts, $S_{i,j}$ is the number of signal counts, and $N_{i,j}$ is the number of observed counts. The index $i$ counts over analysis bins, or fHit bins. The index $j$ counts over the spatial pixels. We use a $14^{\circ}\times14^{\circ}$ ROI where each pixel is $0.1^{\circ}\times0.1^{\circ}$. 

The number of background counts is calculated using a process called `direct integration'~\cite{2012ApJ...750...63A,Abeysekara:2017mjj}, which depends on the approximation that the HAWC data is dominated by background.  In direct integration, the all sky event rate is convolved with an approximation of the local detector efficiency. The local detector efficiency is approximated by counting the events that arrive from the same location in the HAWC field of view as the ROI (e.g. same declination and local hour angle). This convolution is performed over 2 hours. The number of signal counts is calculated by convolving Eq.\ref{eq:dmfluxA} (or Eq.\ref{eq:dmfluxD}) with the HAWC gamma-ray instrument response. Therefore $\langle \sigma v\rangle$ and $\tau_{DM}$ are free in the fit for DM annihilation and decay respectively.

We then calculate a test statistic (TS) to compare the fit with signal to the background-only fit. 
\begin{equation}
TS = -2 \rm{ln} \left(\frac{\mathscr{L}_{0}}{\mathscr{L}^{max}}\right)
\end{equation}
where $\mathscr{L}_{0}$ is the likelihood from the background-only fit and $\mathscr{L}^{max}$ is the likelihood from the best fit with the signal model. A significant detection would have TS $\geq25$. When no significant detection is made we will set 95\% confidence level (CL) upper limit (UL) on $\langle \sigma v\rangle$ and lower limit (LL) on $\tau_{DM}$ as the values where the $TS$ increases to 2.71 relative to the best fit value~\citep{Agashe:2014kda,Rolke:2004mj}.

Very high energy photons ($E>100$TeV) are expected to attenuate on the extragalactic background light~\cite{Venters:2010bq} and even the Milky Way interstellar Radiation Field~\cite{Moskalenko:2005ng}. Our analysis uses photons up to $\sim$50 TeV where these effects are negligible ($<$10\%).

%% file: results.tex
\section{Results}\label{sec:results}

\begin{figure}[ht]
    \centering
    \includegraphics[width=0.48\textwidth]{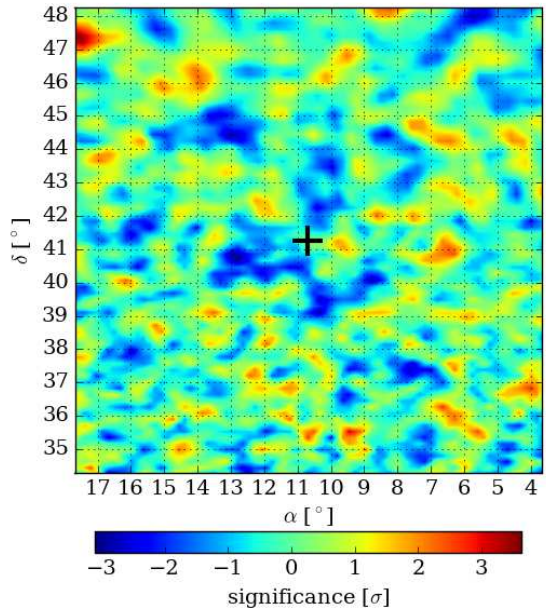}
    \caption{Significance map from analysis bins 1-9 in the $14^{\circ}\times14^{\circ}$ ROI. Mercator projection is used.}\label{fig:map}
  \end{figure}

We searched using the MIN, MED, and MAX DM halo models for DM masses 1, 2.5, 5, 10, 25, 50, 100 TeV for DM annihilating or decaying into $b\bar{b}$, $t\bar{t}$, $\tau^{+}\tau^{-}$, $\mu^{+}\mu^{-}$, and $W^{+}W^{-}$. No significant gamma-ray excess was found in any of the fits and therefore we set limits. Note these limits are calculated using the prompt gamma-ray emission only (no secondary inverse Compton gamma rays included). Figure~\ref{fig:map} is the significance map of the $14^{\circ}\times14^{\circ}$ ROI along with contours for the MED DM halo model $J-$factor.

\begin{figure}[ht]
    \centering
    \includegraphics[width=0.48\textwidth]{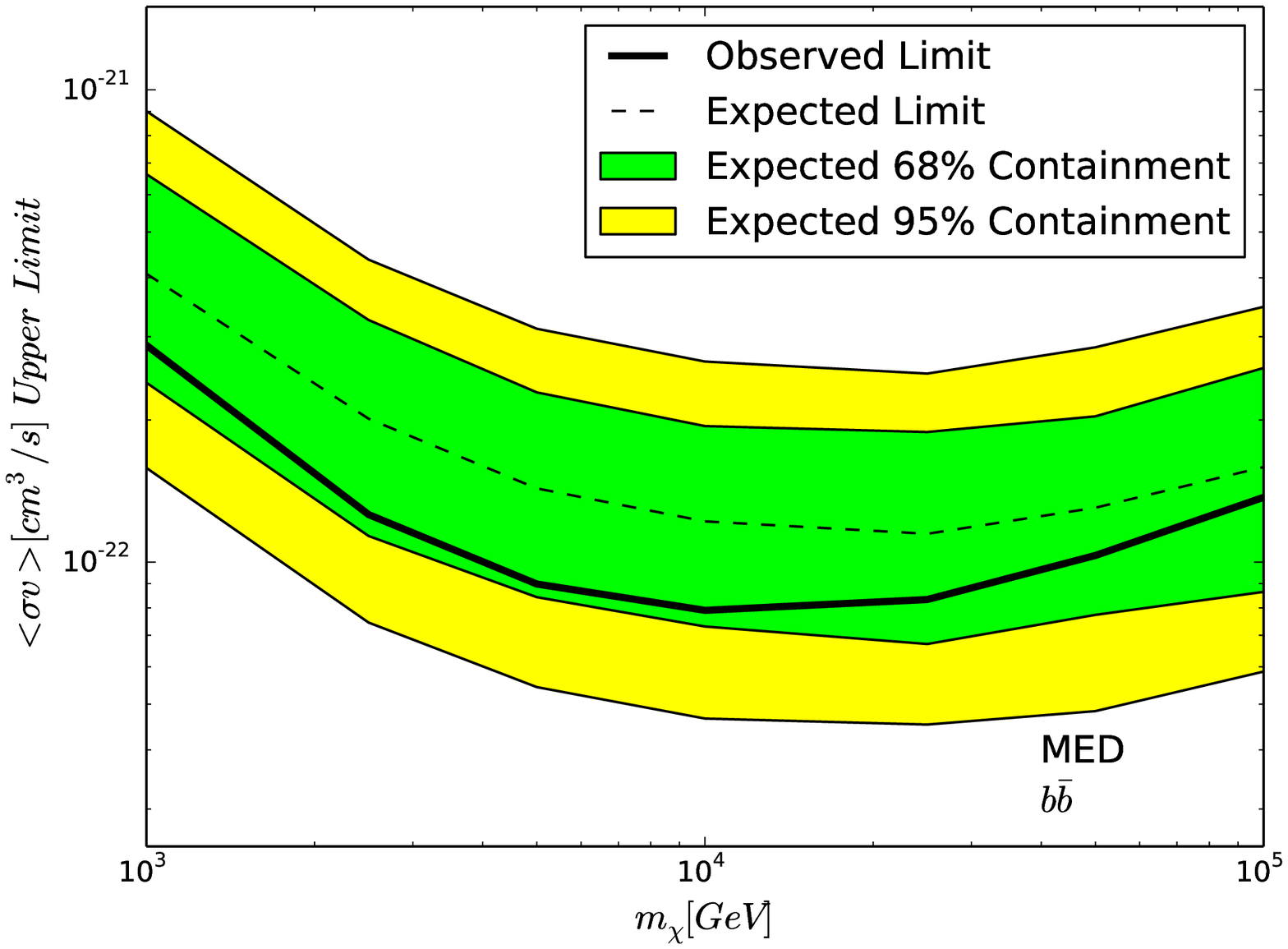}
    \includegraphics[width=0.48\textwidth]{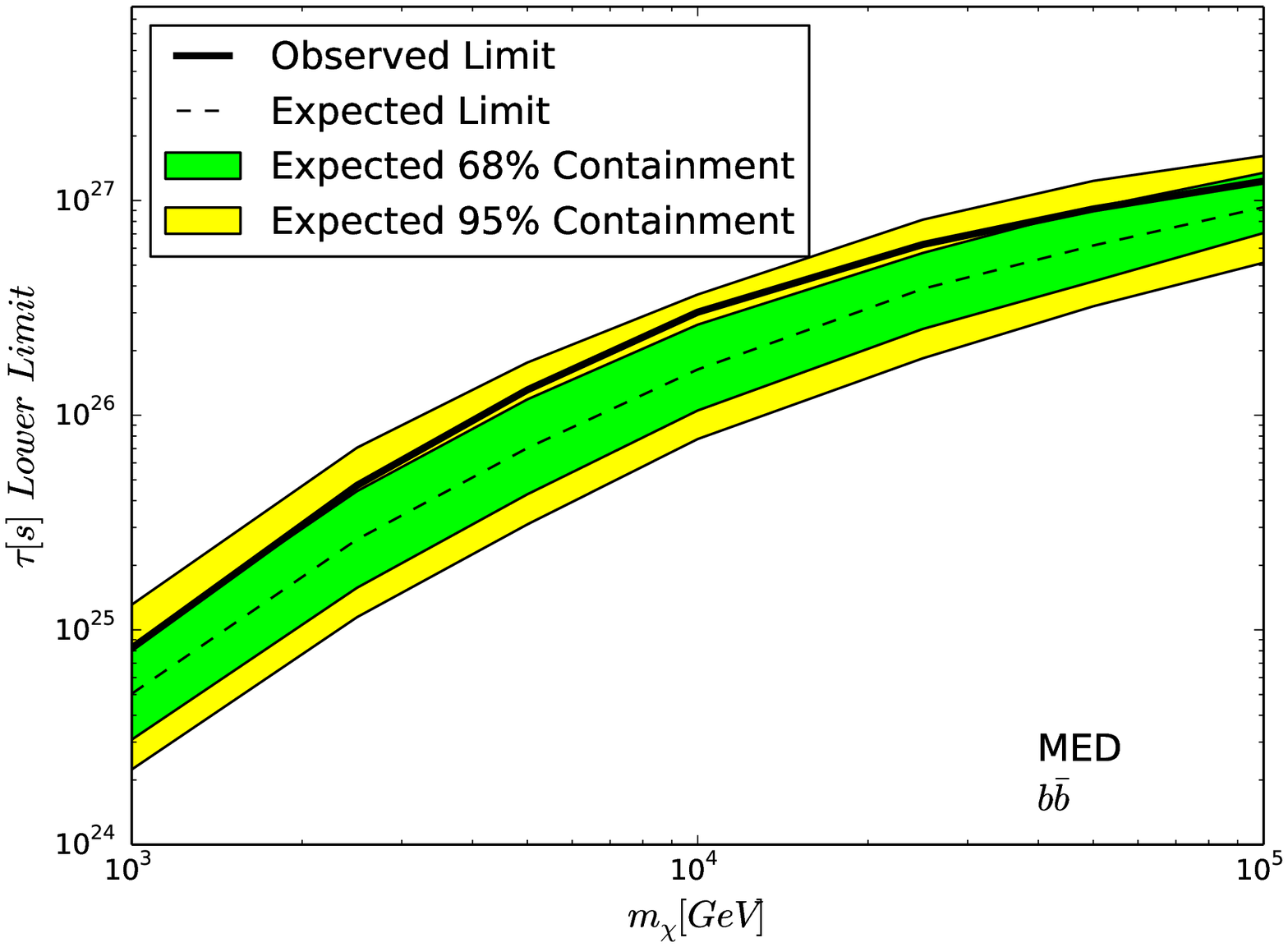} 
    \caption{DM $\langle \sigma v\rangle$ and $\tau_{DM}$ limits for annihilation (top) and decay (bottom) respectively into $b\bar{b}$ for our benchmark DM halo model (MED). Also shown are the expected limits, 68\% containment, and 95\% containment for the null hypothesis based on the HAWC sensitivity.\label{fig:brazil}}
  \end{figure}

Figure \ref{fig:brazil} shows the observed limits for DM annihilation and decay to $b\bar{b}$ for our benchmark DM halo model (MED). The expected limits, 68\% and 95\% containment for the null hypothesis are also shown. The containment bands and expected limit are calculated using 1000 simulations with no DM. Note that the bands are purely statistical. From this we can see that our observed limits are about a $1\sigma$ downward fluctuation for most masses.

\begin{figure}[ht]
    \centering
    \includegraphics[width=0.48\textwidth]{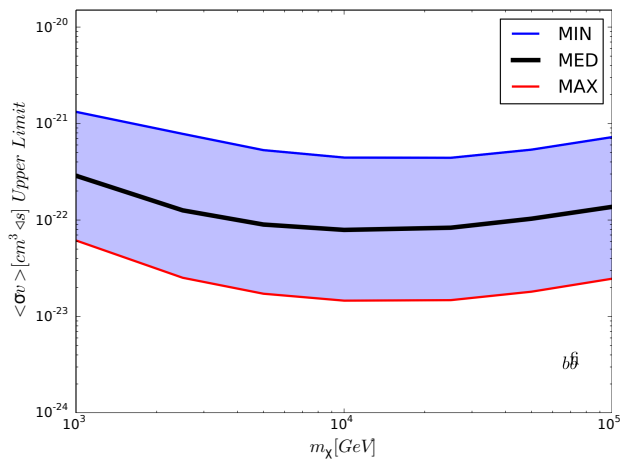}
    \caption{DM $\langle \sigma v\rangle$ 95\% confidence level upper limits for the $b\bar{b}$ channel and all three DM halo models. Results from the $t\bar{t}$, $\tau^{+}\tau^{-}$, $\mu^{+}\mu^{-}$, and $W^{+}W^{-}$ channels can be found in App~\ref{sec:app}. \label{fig:allJAnn}}
  \end{figure}

\begin{figure}[ht]
    \centering
    \includegraphics[width=0.48\textwidth]{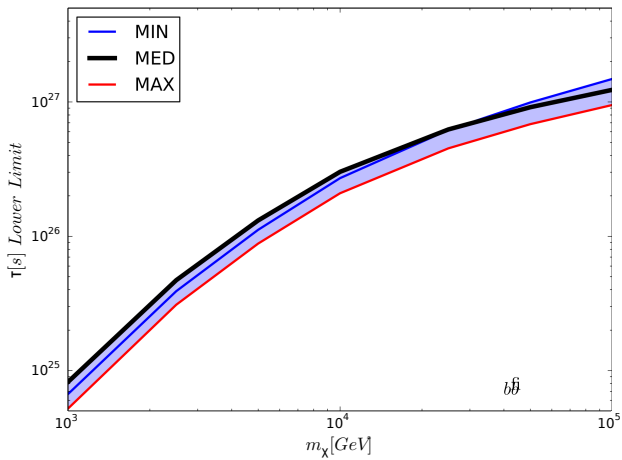}
    \caption{DM $\tau_{DM}$ 95\% confidence level lower limits for the $b\bar{b}$ channel and all three DM halo models. Results from the $t\bar{t}$, $\tau^{+}\tau^{-}$, $\mu^{+}\mu^{-}$, and $W^{+}W^{-}$ channels can be found in App~\ref{sec:app}. \label{fig:allJDec}}
  \end{figure}

The limits from the MIN, MED, and MAX DM halo models for the $b\bar{b}$ channel are shown in Figs.~\ref{fig:allJAnn} and \ref{fig:allJDec}. The limits from the $t\bar{t}$, $\tau^{+}\tau^{-}$, $\mu^{+}\mu^{-}$, and $W^{+}W^{-}$ channels are shown in App~\ref{sec:app}. Results from all models tested are listed in Tab.~\ref{tab:limAnn} and Tab.~\ref{tab:limDec}. For each channel, the benchmark DM halo model is shown in black along with the MIN and MAX models to show the uncertainty in the limits due to DM halo modeling. The difference between the MIN and MAX scenarios is larger for annihilation because the main differences in the DM density between the two models come from the central DM density of M31 and the contribution from subhalos. Since the J-factor is calculated using the square of the DM density, therefore these central differences produce larger differences in the J-factor than the D-factor.

These limits on the DM annihilation cross section and lifetime depend on the modeling parameters chosen (e.g. MED, DM DM$\rightarrow b\bar{b}$). However, these limits are ultimately determined by a lack of gamma-ray flux detected from M31. Gamma-ray flux limits are more general than those from a specific DM model. Calculating the exact gamma-ray flux limits at specific energies is not possible with the current HAWC data analysis. Instead of binning the events by their energy, we perform the likelihood analysis in analysis bins 1-9. Future HAWC analyses will be able to provide a reconstructed energy for each shower using a multi-parameter characterization including the number of PMTs hit, the zenith angle, and the core location~\citep{Marinelli:2017vzu}.

However, we can derive quasi-differential flux limits using the procedure outlined in Ref.~\cite{Aartsen:2017snx} and Ref.~\cite{Albert:2017vtb}. The quasi-differential flux upper limits were calculated using the analysis bin in small energy ranges. We use 0.5 $log(E/TeV)$ bins and assume a power law function with index 2 as an approximation of the signal.

\begin{equation}
\frac{dF}{dE} = K\left(\frac{E}{TeV}\right)^{-2}.
\label{eq.pl}
\end{equation}

\begin{figure}[h]
    \centering
    \includegraphics[width=0.48\textwidth]{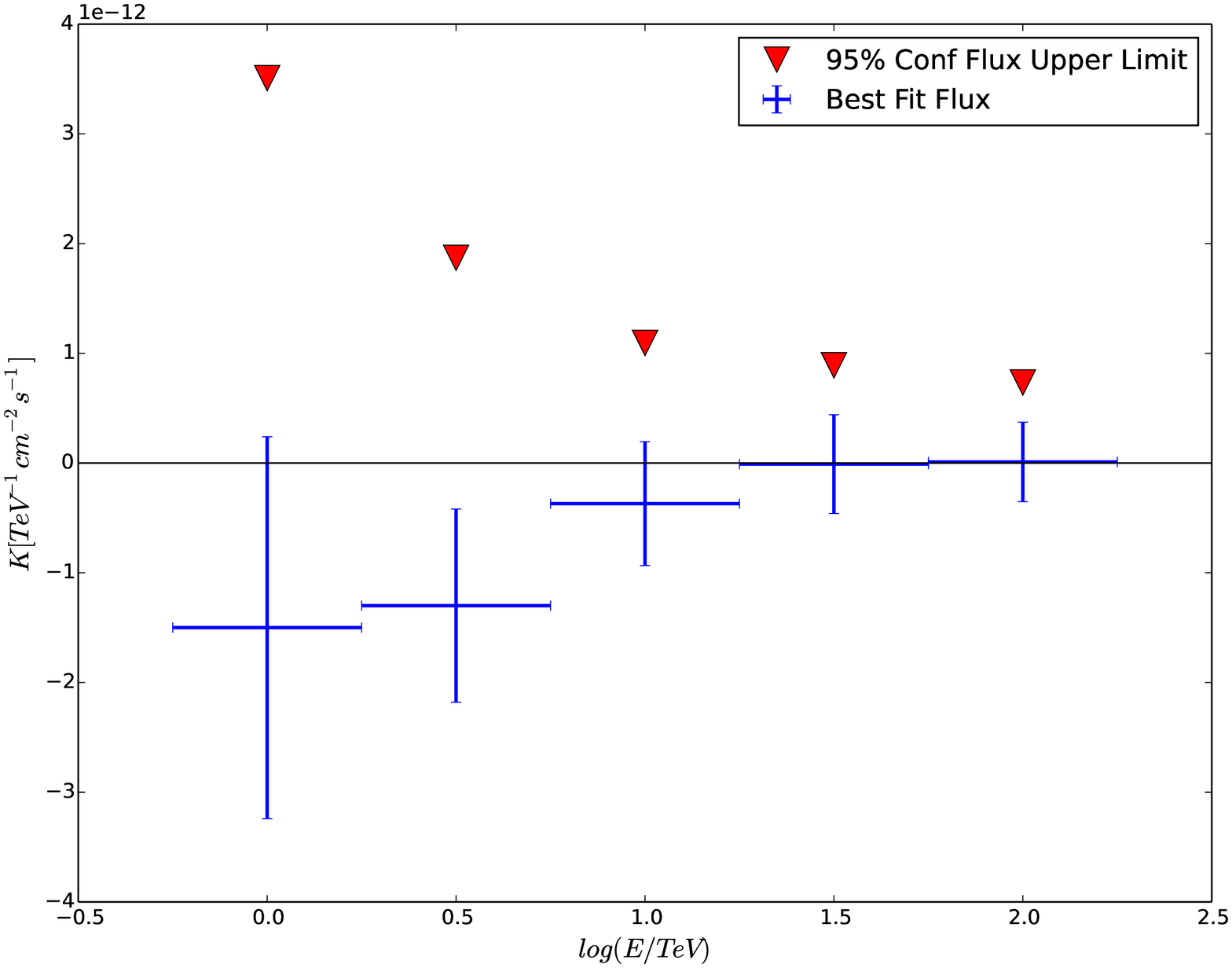}
    \includegraphics[width=0.48\textwidth]{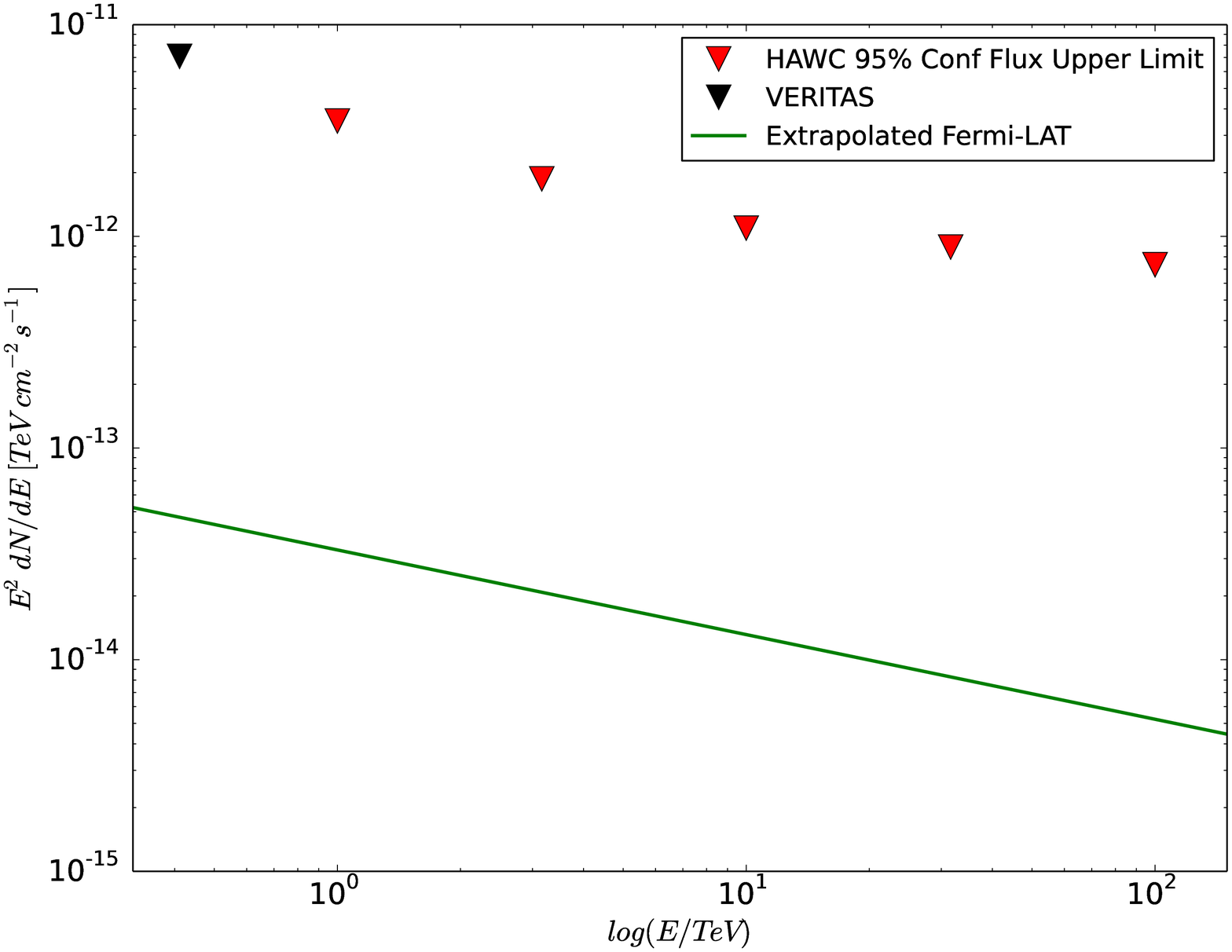}
    \caption{Quasi-differential gamma-ray flux results for M31. (top) Best fit normalization of a powerlaw with index 2 in 0.5 $log(E/TeV)$ bins is shown in blue. The 95\% confidence level upper limits on the normalization are shown in red triangles. (bottom) The 95\% confidence level upper limits from this analysis and a VERITAS analysis of M31~\cite{Bird:2015npa}. Also shown is an extrapolation of the recent best fit to the \it{Fermi}~\rm{LAT observations of M31~\cite{M31Fermi}}. \label{fig:flux}}
  \end{figure}

We find the best fit value of the normalization ($K$) in each bin, allowing $K$ to go negative to account for deficits. We also calculate the 95\% confidence level upper limit by finding the value of $K$ that increases the loglikelihood by 2.71 relative to the best fit value~\citep{Agashe:2014kda,Rolke:2004mj}.  The resulting best fit normalization and 95\% confidence level upper limit is shown in Fig.~\ref{fig:flux}. We find a deficit of gamma rays from 1 - 10 TeV. This is expected since the DM limits show a $\sim 1 \sigma$ deficit (see Fig \ref{fig:brazil}). It should also be noted that the flux upper limit results obtained by this method are similar when modifying Eq.~\ref{eq.pl} to have indices ranging from 0 to 3 at the center of each energy bin~\citep{Aartsen:2017snx}.

%% file: conclusions.tex
\section{Discussion and Conclusions}\label{sec:conclusion}

We searched for gamma rays from DM annihilation and decay in M31 and did not find any significant detection. The limits on the DM annihilation cross section and DM decay lifetime are given in Section \ref{sec:results} and App. \ref{sec:app}. 

\begin{figure}[ht]
    \centering
    \includegraphics[width=0.48\textwidth]{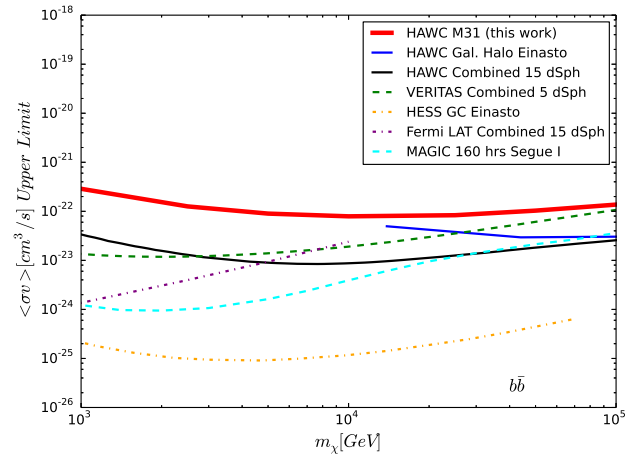}
    \caption{DM $\langle \sigma v\rangle$ 95\% confidence level upper limits for our benchmark DM halo model (MED,red) compared with recent DM searches with HAWC and other gamma-ray experiments when available for the $b\bar{b}$ channel. Results from the $t\bar{t}$, $\tau^{+}\tau^{-}$, $\mu^{+}\mu^{-}$, and $W^{+}W^{-}$ channels can be found in App~\ref{sec:app}. The limits obtained from the HAWC Galactic Halo search (specifically the northern \textit{Fermi} Bubble region)~\citep{harding} are shown in blue.  The HAWC limits obtained from a joint analysis of 15 dwarf spheroidal galaxies~\citep{Albert:2017vtb} are shown in black. The VERITAS limits obtained from a joint analysis of 5 dwarf spheroidal galaxies~\citep{Archambault:2017wyh} are shown in green. The \textit{Fermi} Large Area Telescope limits obtained from a joint analysis of 15 dwarf spheroidal galaxies~\citep{Ackermann:2015zua} are shown in purple. The MAGIC limits obtained from 160 hours of observation of the dwarf spheroidal galaxy Segue I~\citep{Aleksic:2013xea} are shown in cyan. Finally the H.E.S.S. limits from observations of the Galactic Center~\citep{Abdallah:2016ygi} are shown in orange. \label{fig:compAnn}}
 \end{figure}

\begin{figure}[ht]
    \centering
    \includegraphics[width=0.48\textwidth]{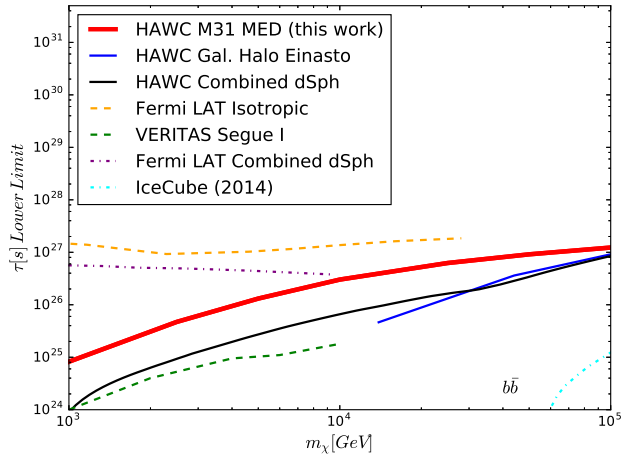}
    \caption{DM $\tau_{DM}$ 95\% confidence level lower limits for our benchmark DM halo model (MED, red) compared with recent DM searches with HAWC and other gamma-ray experiments when available for the $b\bar{b}$ channel. Results from the $t\bar{t}$, $\tau^{+}\tau^{-}$, $\mu^{+}\mu^{-}$, and $W^{+}W^{-}$ channels can be found in App~\ref{sec:app}. The limits obtained from the HAWC Galactic Halo search (specifically the northern \textit{Fermi} Bubble region)~\citep{harding} are shown in blue.  The HAWC limits obtained from a joint analysis of 15 dwarf spheroidal galaxies~\citep{Albert:2017vtb} are shown in black. The VERITAS limits obtained from 48 hours of observation on the dwarf spheroidal galaxy Segue I~\citep{Aliu:2012ga} are shown in green. The limits using \textit{Fermi} LAT data of 19 dwarf spheroidal galaxies~\citep{Baring:2015sza} are shown in purple.\label{fig:compDec}. Limits from an analysis of the isotropic extragalactic background using \textit{Fermi} LAT data is shown in orange~\cite{Esmaili:2015xpa}. Decay limits from an analysis of neutrinos by IceCube are shown in cyan~\cite{Esmaili:2014rma}.}
\end{figure}

We compare our benchmark model (MED) to other recent HAWC analyses in Fig~\ref{fig:compAnn} and \ref{fig:compDec}. Specifically we compare to a combined analysis of 15 dwarf spheroidal galaxies~\citep{Albert:2017vtb} and also limits obtained by studying the northern \textit{Fermi} bubble region assuming the Einasto profile~\cite{harding}. Our annihilation limits are less constraining than the dwarf spheroidal limits. This is because the dwarf annihilation limits are dominated by Triangulum II, which had a $J$-factor of $2.75\times10^{20} [\rm{GeV}^{2}\rm{cm}^{-5}]$ in that work. That $J-$factor is slightly larger than our MED $J-$factor of $1.27\times10^{20} [\rm{GeV}^{2}\rm{cm}^{-5}]$. Our decay limits are better than the HAWC dwarf limits which are dominated by Coma Berenices with a $D-$factor of $2.09\times10^{19} [\rm{GeV cm}^{-2}]$. Compared to the $D-$factor values from the dwarf analysis, our MED value of $1.56\times10^{20} [\rm{GeV cm}^{-2}]$ is larger since the M31 DM halo mass is larger making it not surprising that the M31 limits are more constraining than the dwarf limits.

We also compare our limits to limits obtained from constraining the gamma-ray flux in the northern \textit{Fermi} bubbles region~\citep{harding}. Though several DM density profiles were considered in that paper, we compare to the Einasto profile, which gave the strongest constraints in that work. Our annihilation and decay limits are more constraining than the Galactic Halo annihilation limits for most masses considered.  It should be noted that the Galactic Halo limits extend to $m_{DM}=10$ PeV.

We also compare our annihilation limits to those obtained by other gamma-ray experiments.  In all channels, our limits nicely complement those from the \textit{Fermi} LAT, VERITAS, and MAGIC.  In all channels the most constraining limits are from H.E.S.S. observations of the Galactic Center~\citep{Abdallah:2016ygi}. This is partially due to the Galactic Center having a larger $J$-factor and the fact that H.E.S.S. performed deep observations of the Galactic Center. It is worth noting that the $J$-factor in the Galactic Center is not well constrained and has large uncertainties. 
%Also, it is important to search for signals in other targets with other instruments. Each DM target represents a discovery space where a DM signal could be found. Additionally, a detection of gamma rays from DM annihilation or decay will need to be seen in multiple targets to truly confirm a DM origin.

We compare our decay limits with the limits from VERITAS observations of Segue I~\citep{Aliu:2012ga}, \textit{Fermi} LAT observations of 19 dwarf spheroidal galaxies~\citep{Baring:2015sza}, an analysis of the extragalactic background light with \textit{Fermi} LAT observations~\cite{Esmaili:2015xpa}, and results from IceCube's search for neutrinos from DM decay~\cite{Esmaili:2014rma}. Additionally we compare to the other HAWC searches. The MED halo model limits obtained in this work are the most constraining for masses from 25 TeV to 100 TeV in the $b\bar{b}$, $t\bar{t}$ and $\mu^{+}\mu^{-}$ channels. 
 
Additionally we derived quasi-differential flux limits from 1 TeV to 100 TeV. Previous high-energy gamma-ray limits on M31 have been derived by VERITAS. They calculated the 95\% confidence level upper flux limit to be  $6.9\times10^{-12}\ \rm{TeV}^{-1}\rm{cm}^{-2} \rm{s}^{-1}$ at 416.9 GeV and $2.7\times10^{-11}\ \rm{TeV}^{-1} \rm{cm}^{-2} \rm{s}^{-1}$ at 346.7 GeV~\citep{Bird:2015npa}. Our limits nicely complement the VERITAS limits since they extend to higher energies. We also note that M31 has not been observed by the neutrino detector IceCube~\citep{Aartsen:2013dxa}.

In conclusion, we searched for gamma rays from M31 from DM annihilation or decay. No gamma-ray excesses were detected and limits were placed on the DM annihilation cross section and decay lifetime. We also present quasi-differential gamma-ray flux limits for M31.  Our annihilation limits complement other searches using HAWC and other gamma-ray observatories. Our decay limits are the most constraining from 25 TeV to 100 TeV in the $b\bar{b}$ and $t\bar{t}$ channels.  Continued observation of M31 by HAWC, along with analysis upgrades, like the inclusion of shower energy estimators, will improve its sensitivity to detecting gamma rays from M31.

%% file: app1.tex
\section{APPENDIX}\label{sec:app}

\begin{figure*}[ht]
    \centering
    \includegraphics[width=0.48\textwidth]{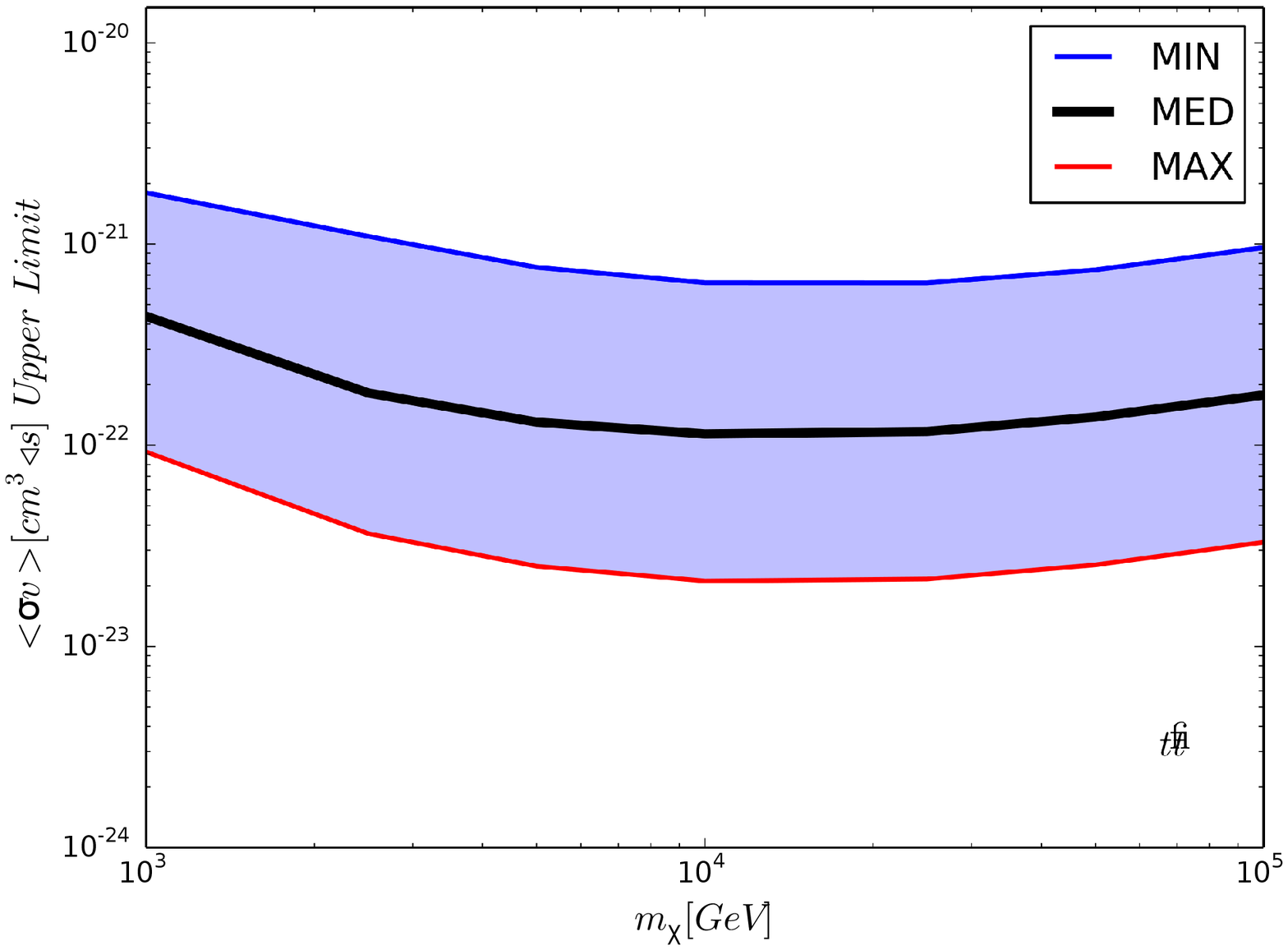}
    \includegraphics[width=0.48\textwidth]{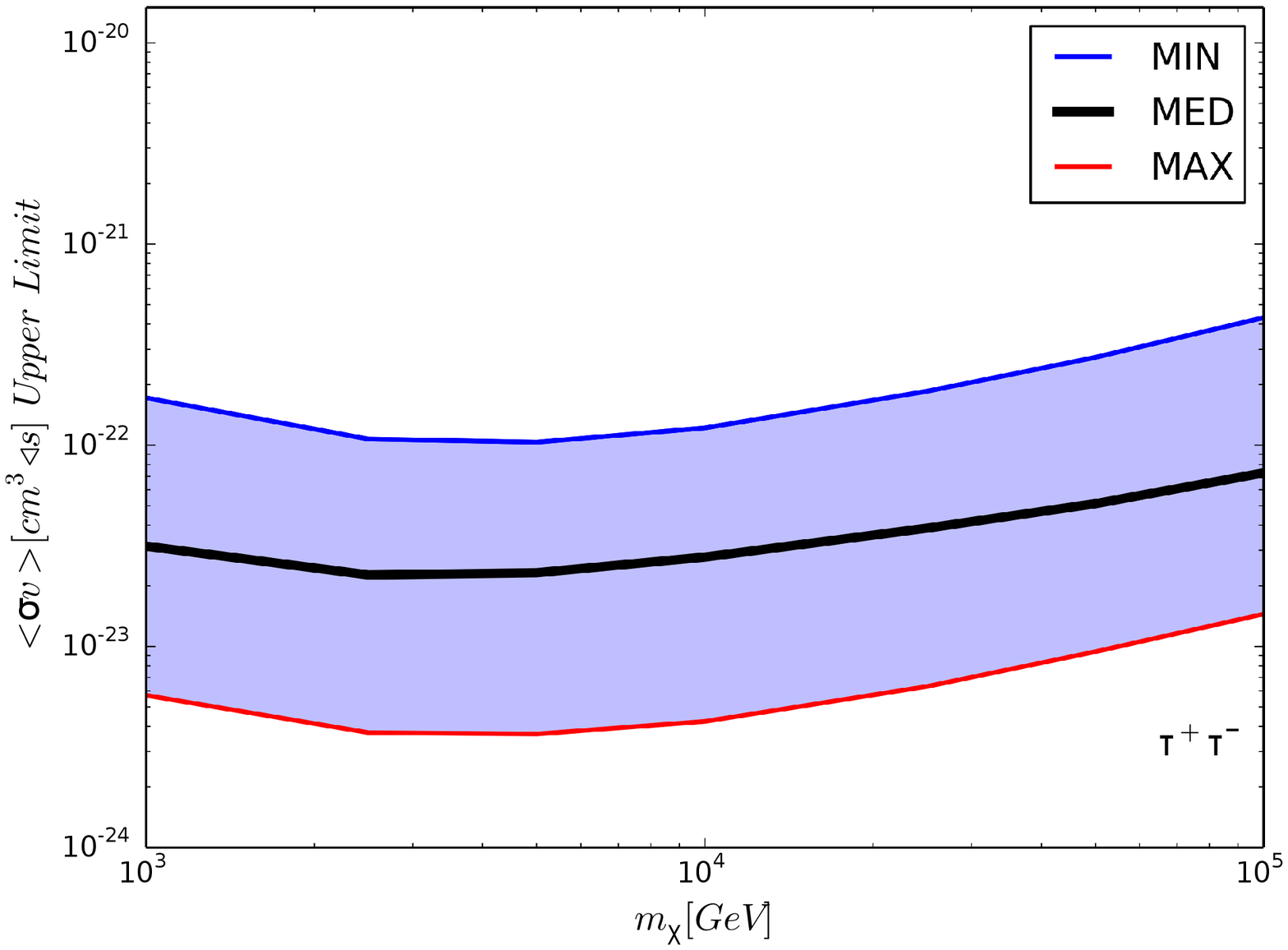}
    \includegraphics[width=0.48\textwidth]{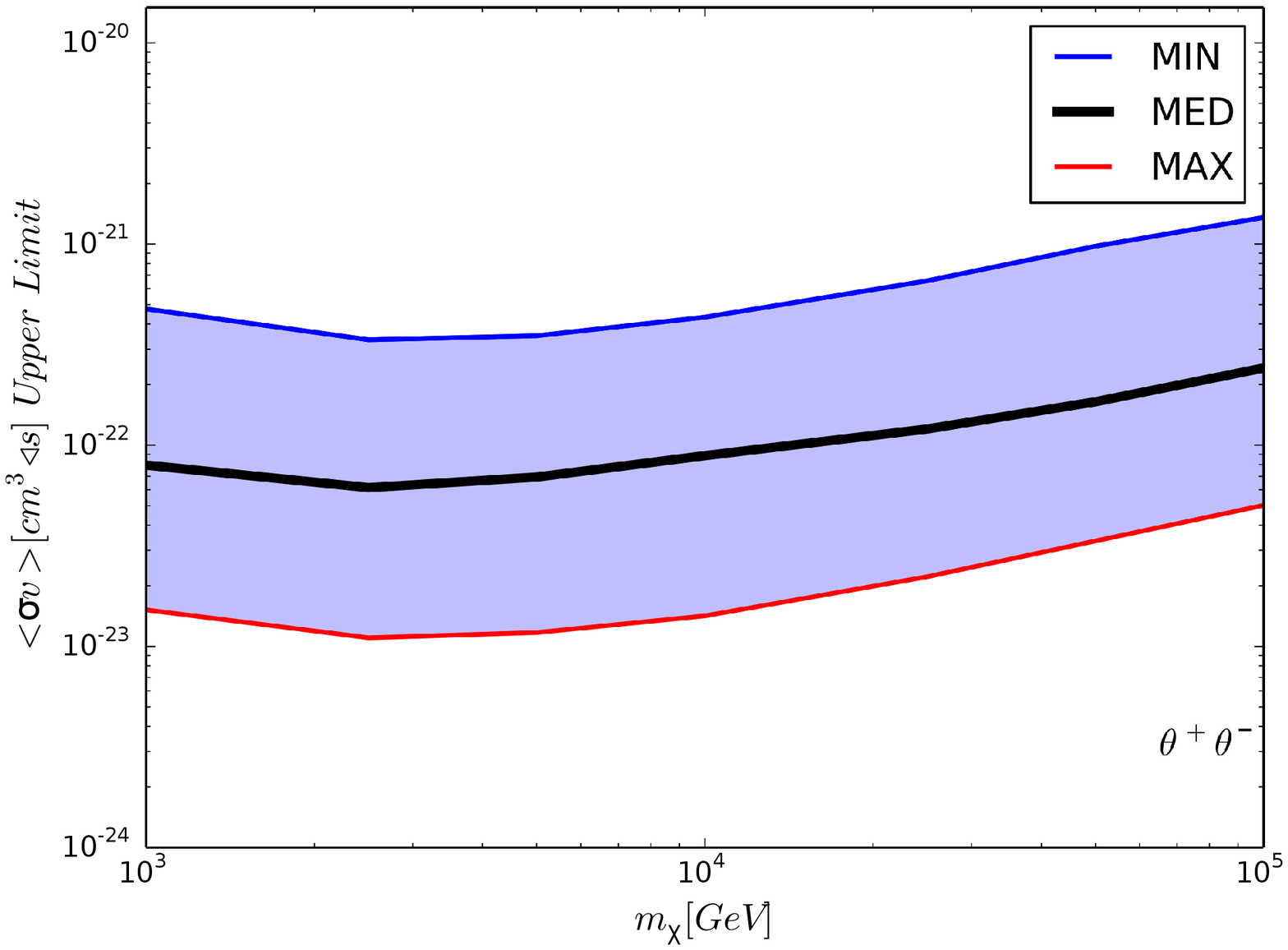}
    \includegraphics[width=0.48\textwidth]{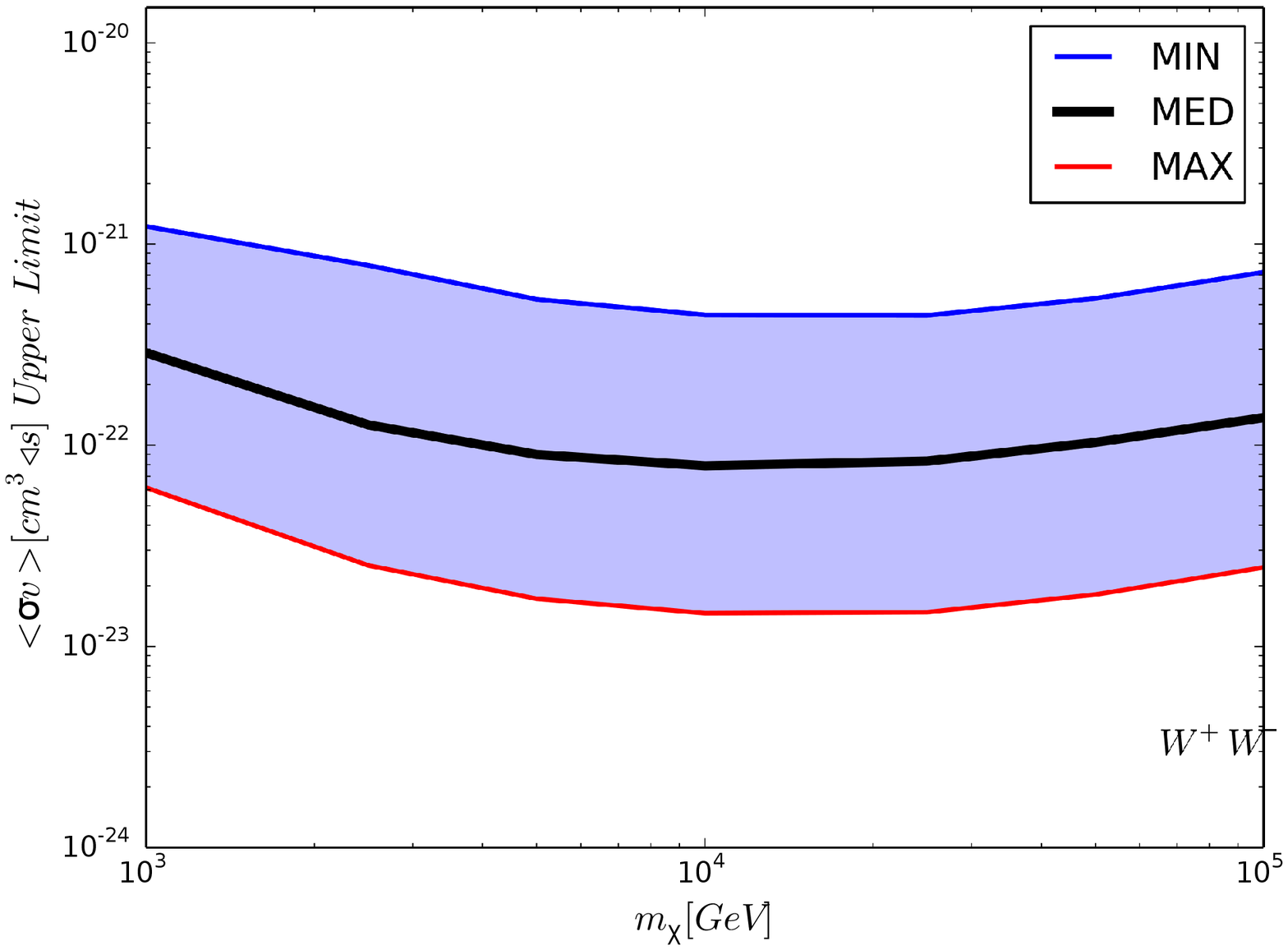}
    \caption{DM $\langle \sigma v\rangle$ 95\% confidence level upper limits for the $t\bar{t}$, $\tau^{+}\tau^{-}$, $\mu^{+}\mu^{-}$, $W^{+}W^{-}$ channels and all three DM halo models.  \label{fig:allJAnn2}}
  \end{figure*}

\begin{figure*}[ht]
    \centering
    \includegraphics[width=0.48\textwidth]{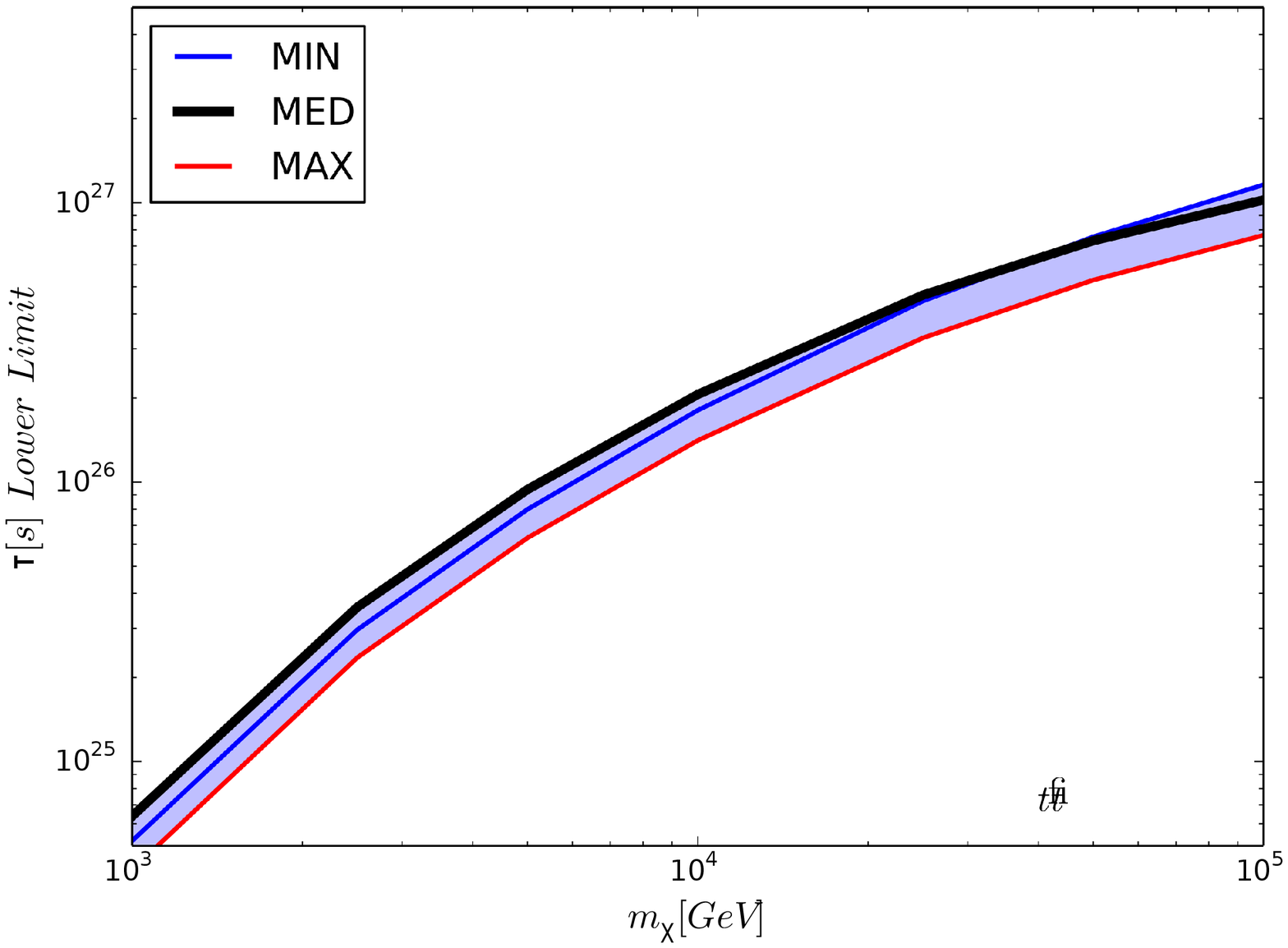}
    \includegraphics[width=0.48\textwidth]{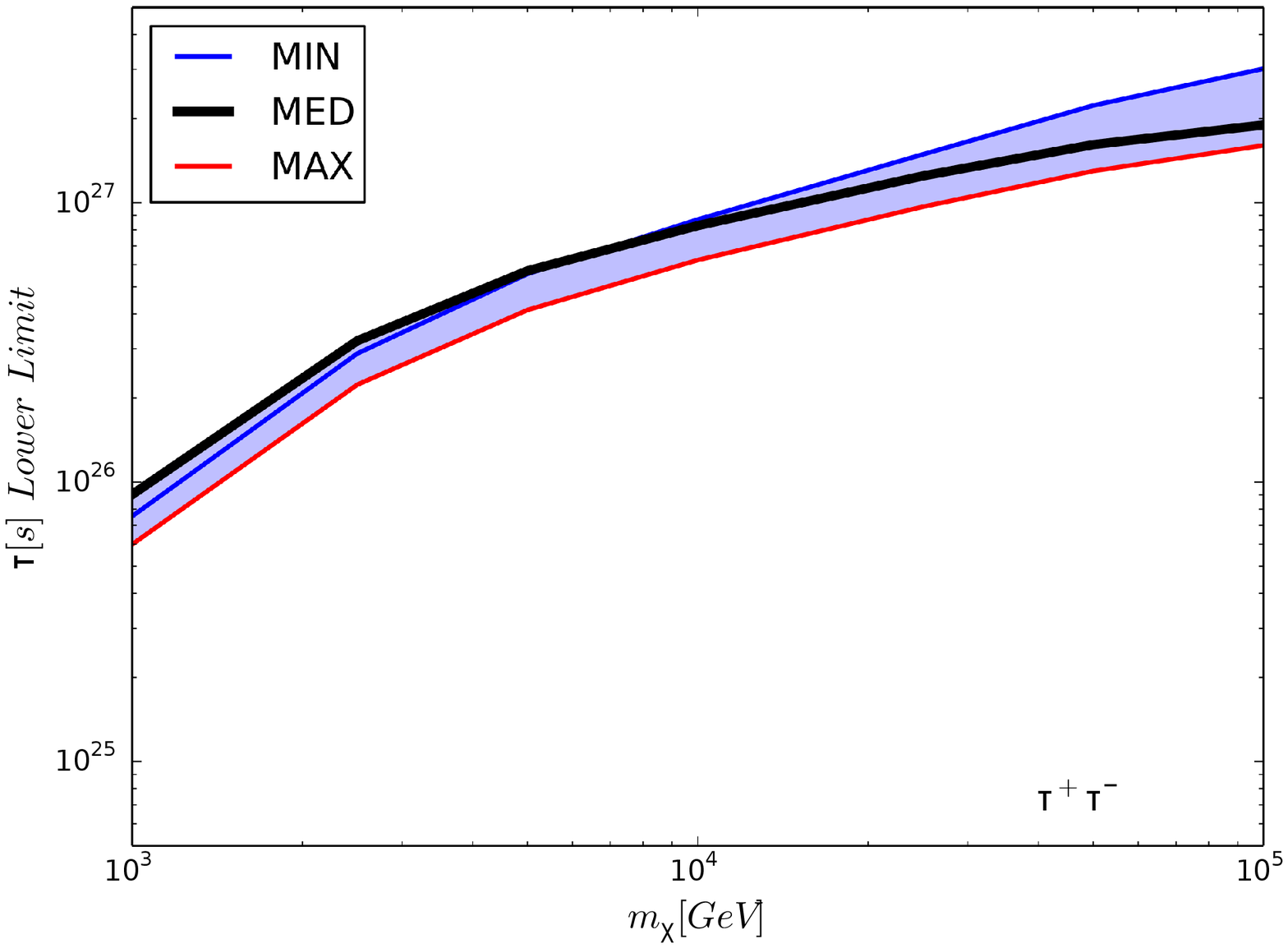}
    \includegraphics[width=0.48\textwidth]{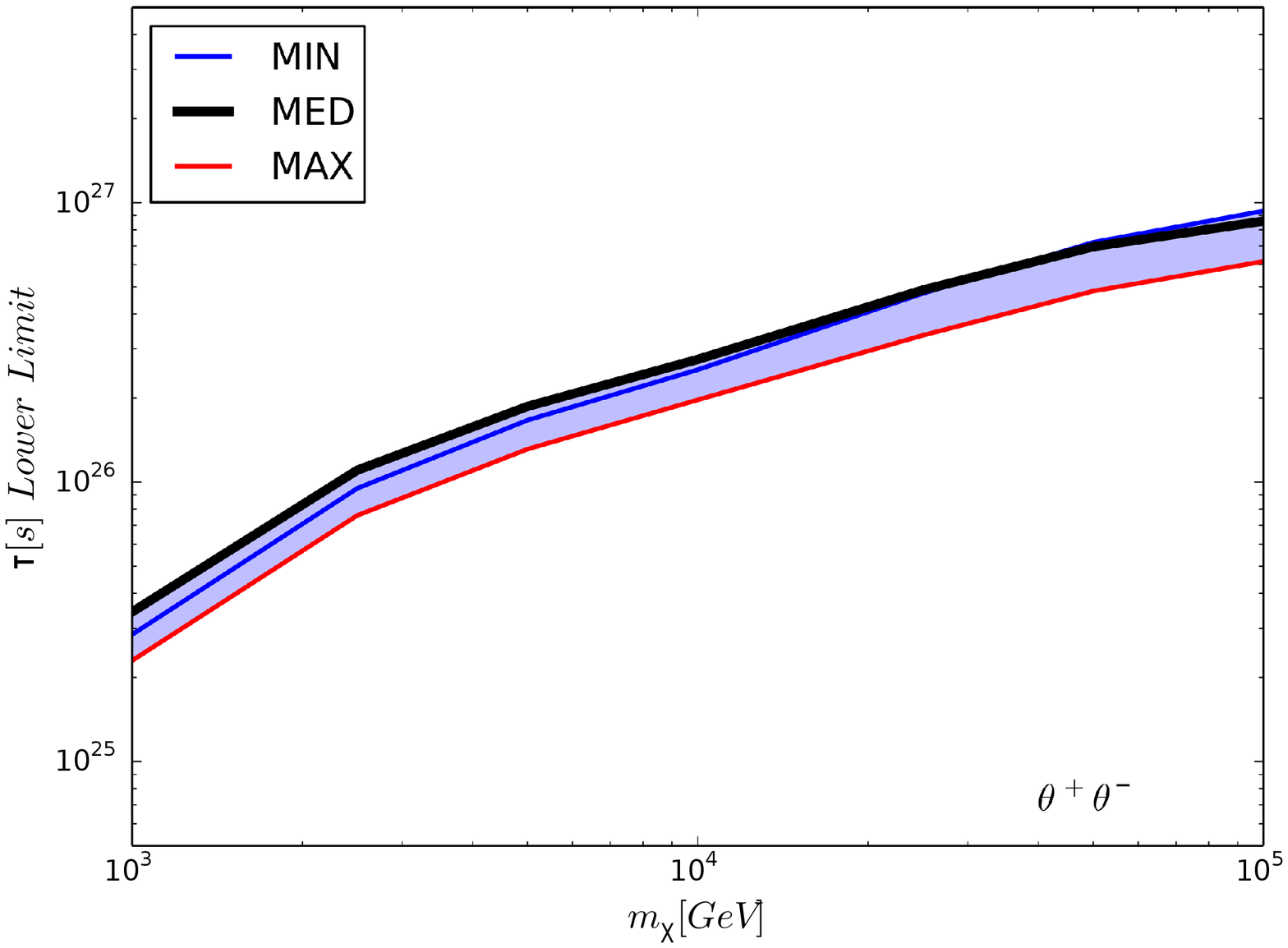}
    \includegraphics[width=0.48\textwidth]{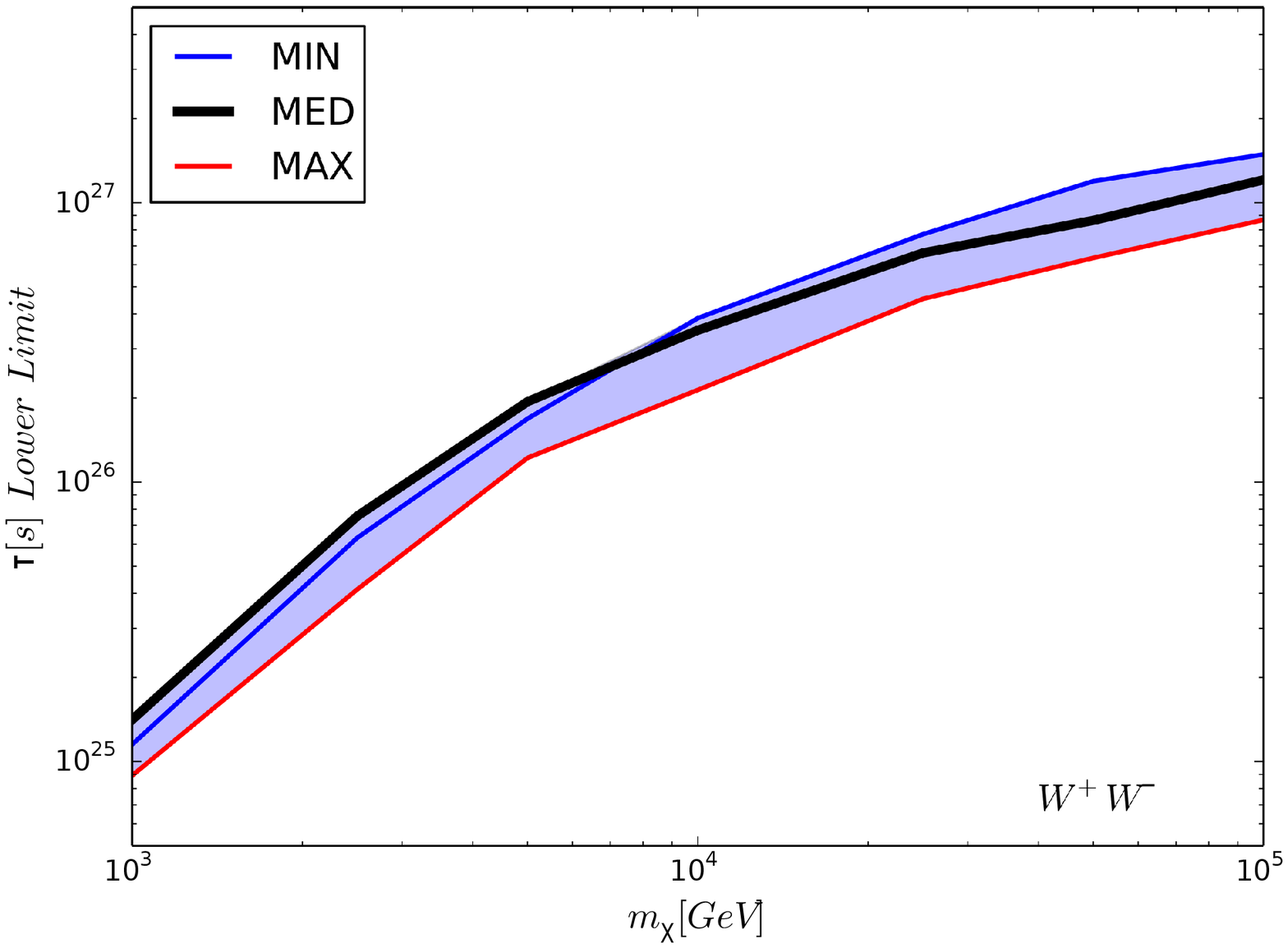}
    \caption{DM $\tau_{DM}$ 95\% confidence level lower limits for the $t\bar{t}$, $\tau^{+}\tau^{-}$, $\mu^{+}\mu^{-}$, $W^{+}W^{-}$ channels and all three DM halo models.  \label{fig:allJDec2}}
  \end{figure*}

\begin{figure*}[ht]
    \centering
    \includegraphics[width=0.48\textwidth]{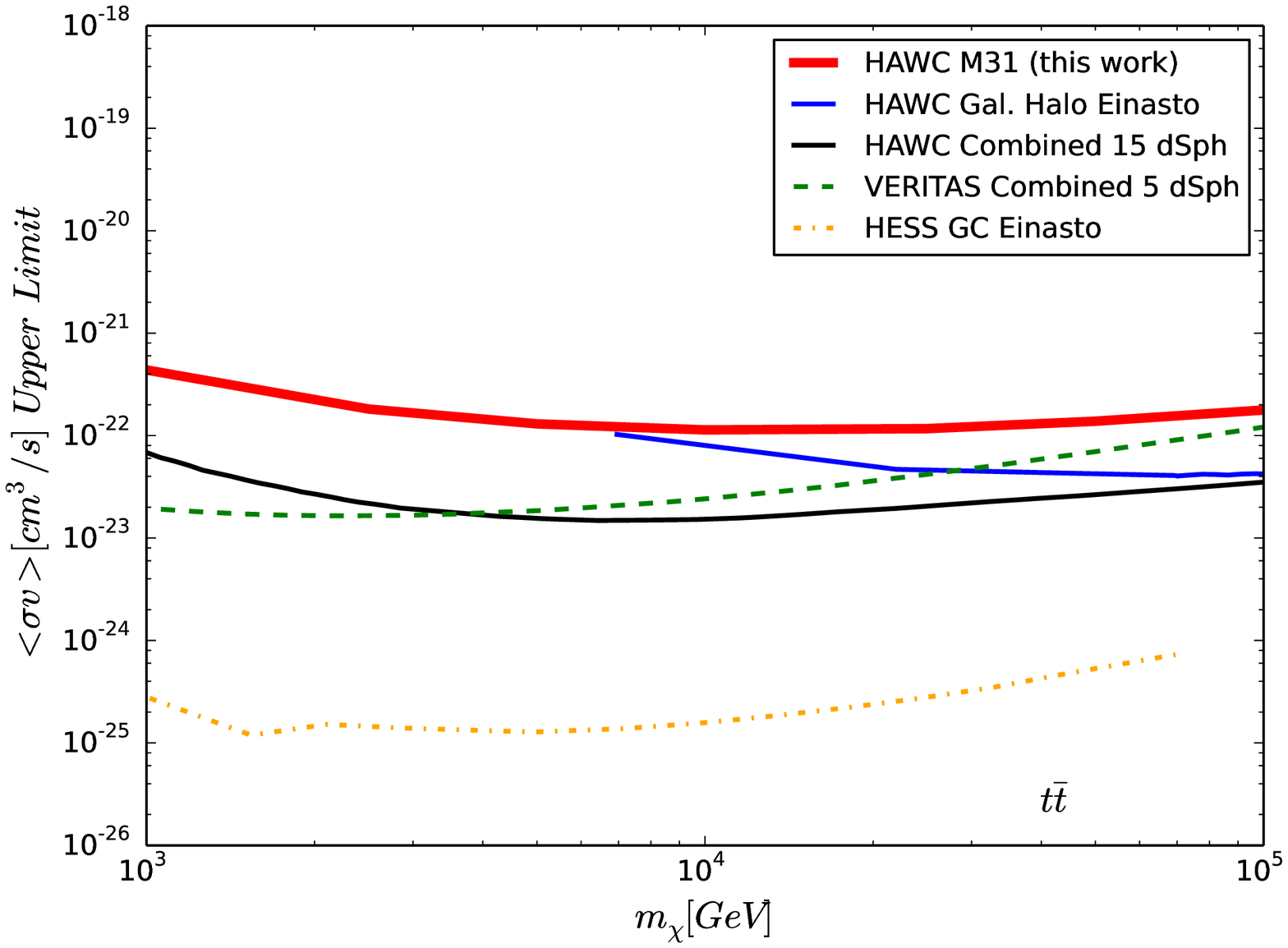}
    \includegraphics[width=0.48\textwidth]{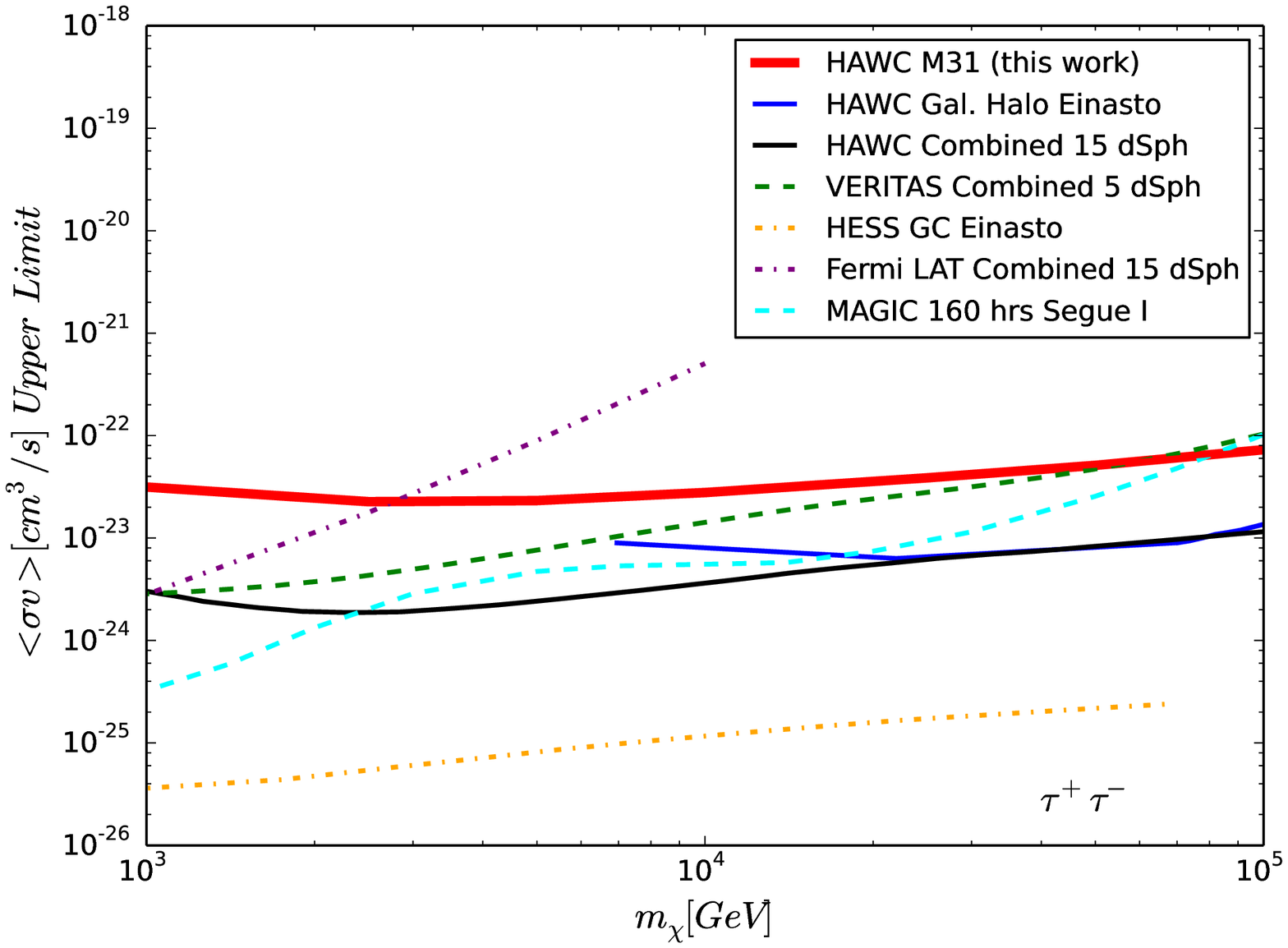}
    \includegraphics[width=0.48\textwidth]{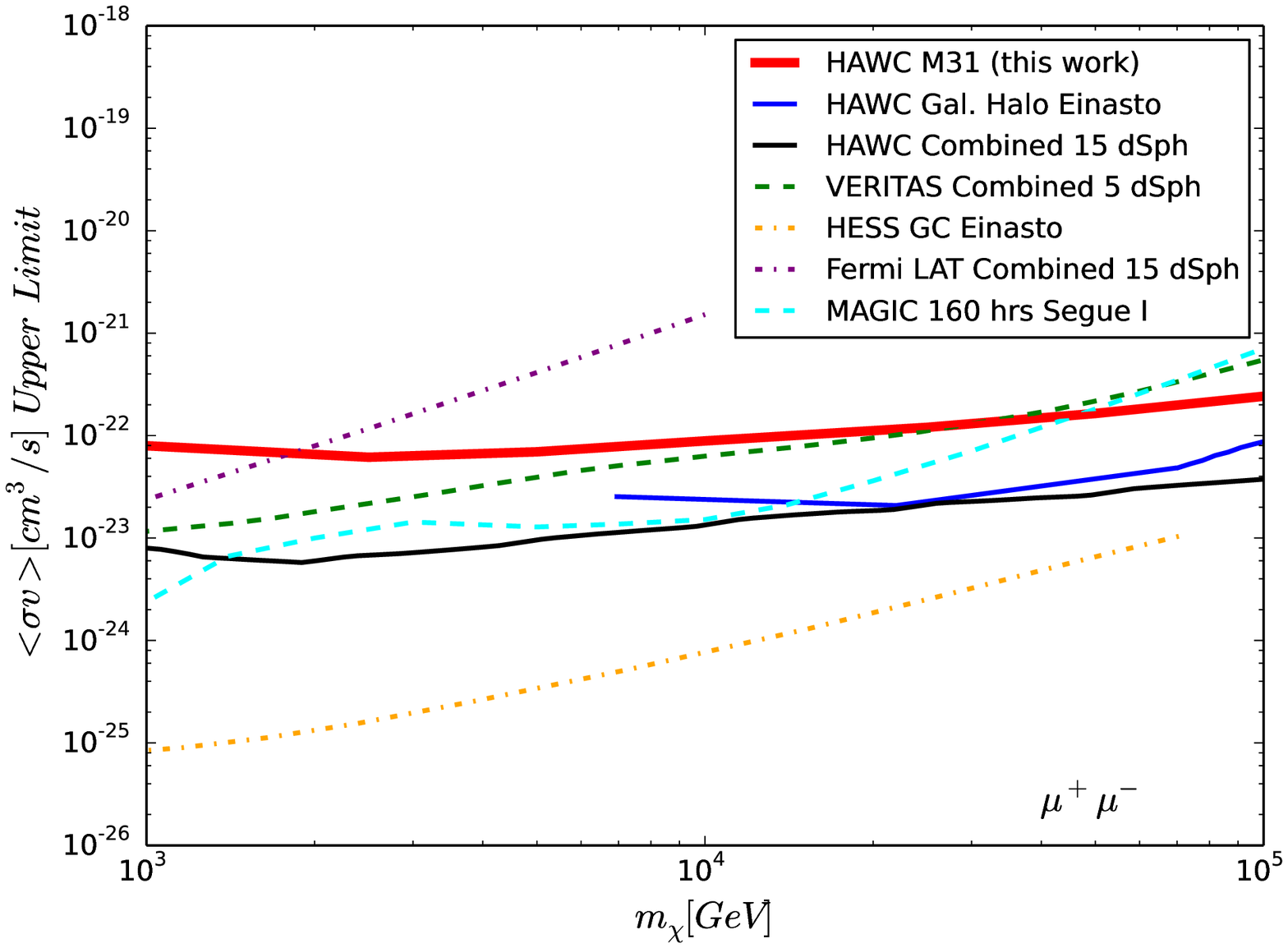}
    \includegraphics[width=0.48\textwidth]{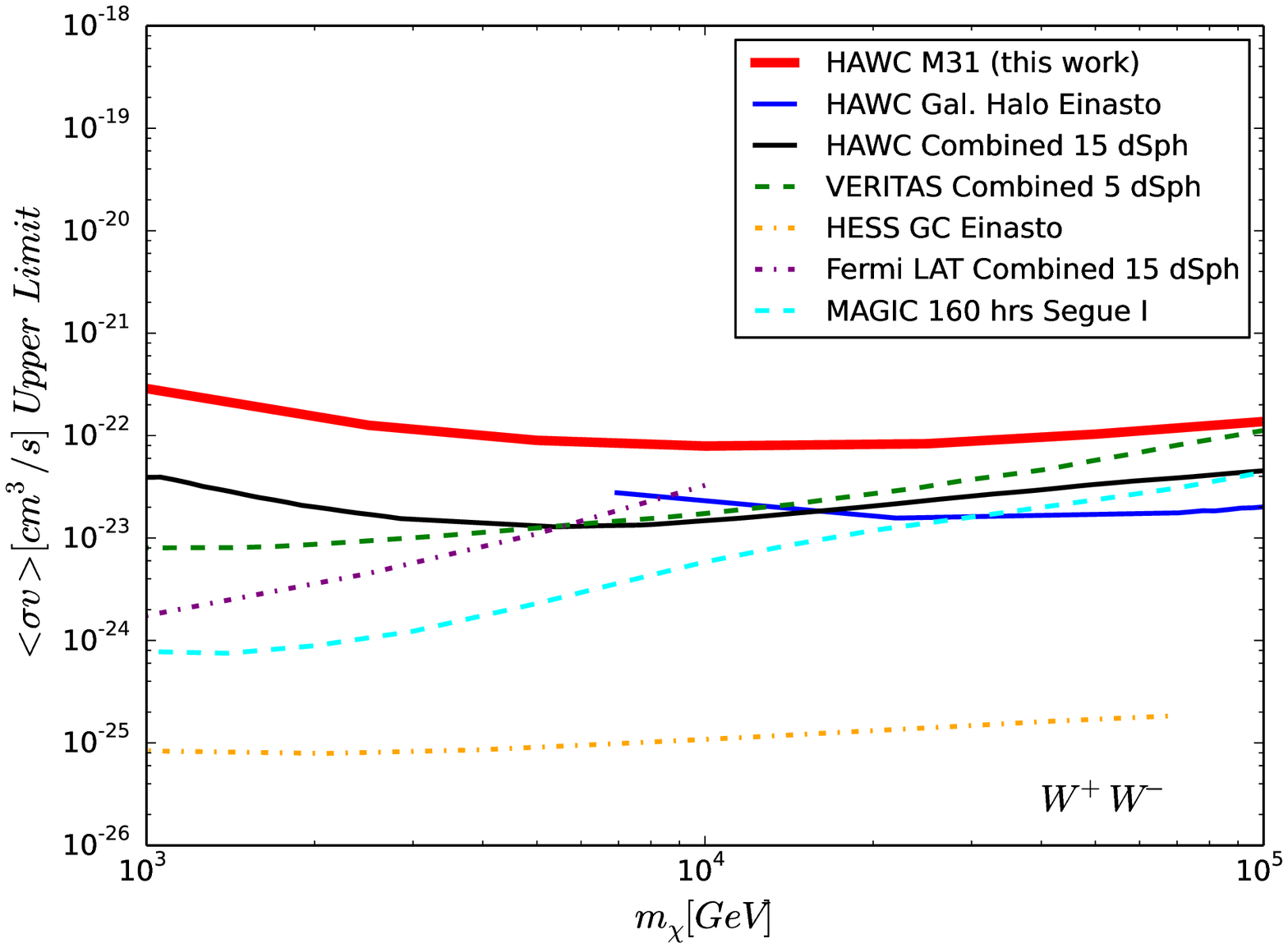}
    \caption{DM $\langle \sigma v\rangle$ 95\% confidence level upper limits for our benchmark DM halo model (MED,red) compared with recent DM searches with HAWC and other gamma-ray experiments when available the $t\bar{t}$, $\tau^{+}\tau^{-}$, $\mu^{+}\mu^{-}$, and $W^{+}W^{-}$ channels. The limits obtained from the HAWC Galactic Halo search (specifically the northern \textit{Fermi} Bubble region)~\citep{harding} are shown in blue.  The HAWC limits obtained from a joint analysis of 15 dwarf spheroidal galaxies~\citep{Albert:2017vtb} are shown in black. The VERITAS limits obtained from a joint analysis of 5 dwarf spheroidal galaxies~\citep{Archambault:2017wyh} are shown in green. The \textit{Fermi} Large Area Telescope limits obtained from a joint analysis of 15 dwarf spheroidal galaxies~\citep{Ackermann:2015zua} are shown in purple. The MAGIC limits obtained from 160 hours of observation of the dwarf spheroidal galaxy Segue I~\citep{Aleksic:2013xea} are shown in cyan. Finally the H.E.S.S. limits from observations of the Galactic Center~\citep{Abdallah:2016ygi} are shown in orange. \label{fig:compAnn2}}
 \end{figure*}

\begin{figure*}[ht]
    \centering
    \includegraphics[width=0.48\textwidth]{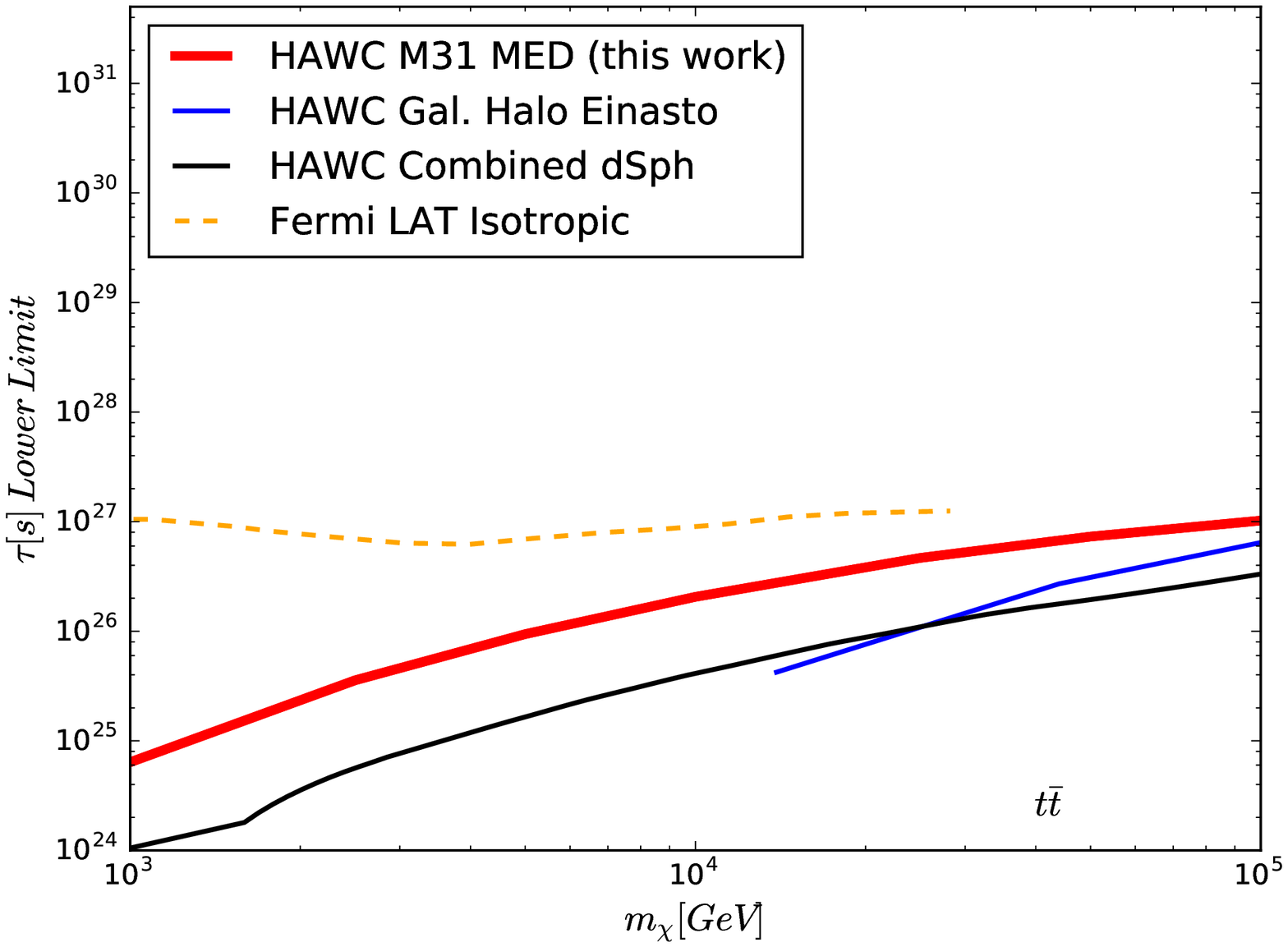}
    \includegraphics[width=0.48\textwidth]{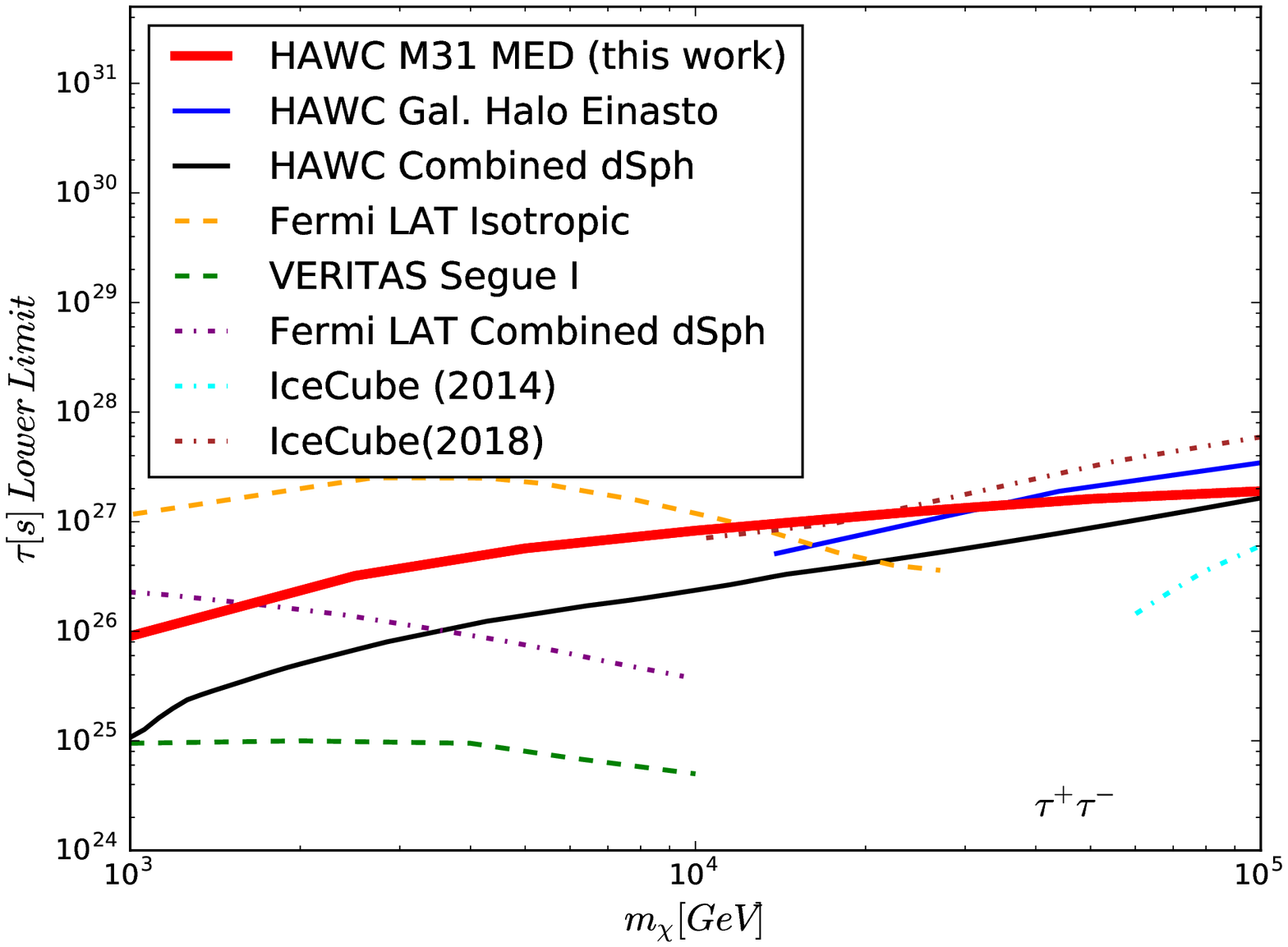}
    \includegraphics[width=0.48\textwidth]{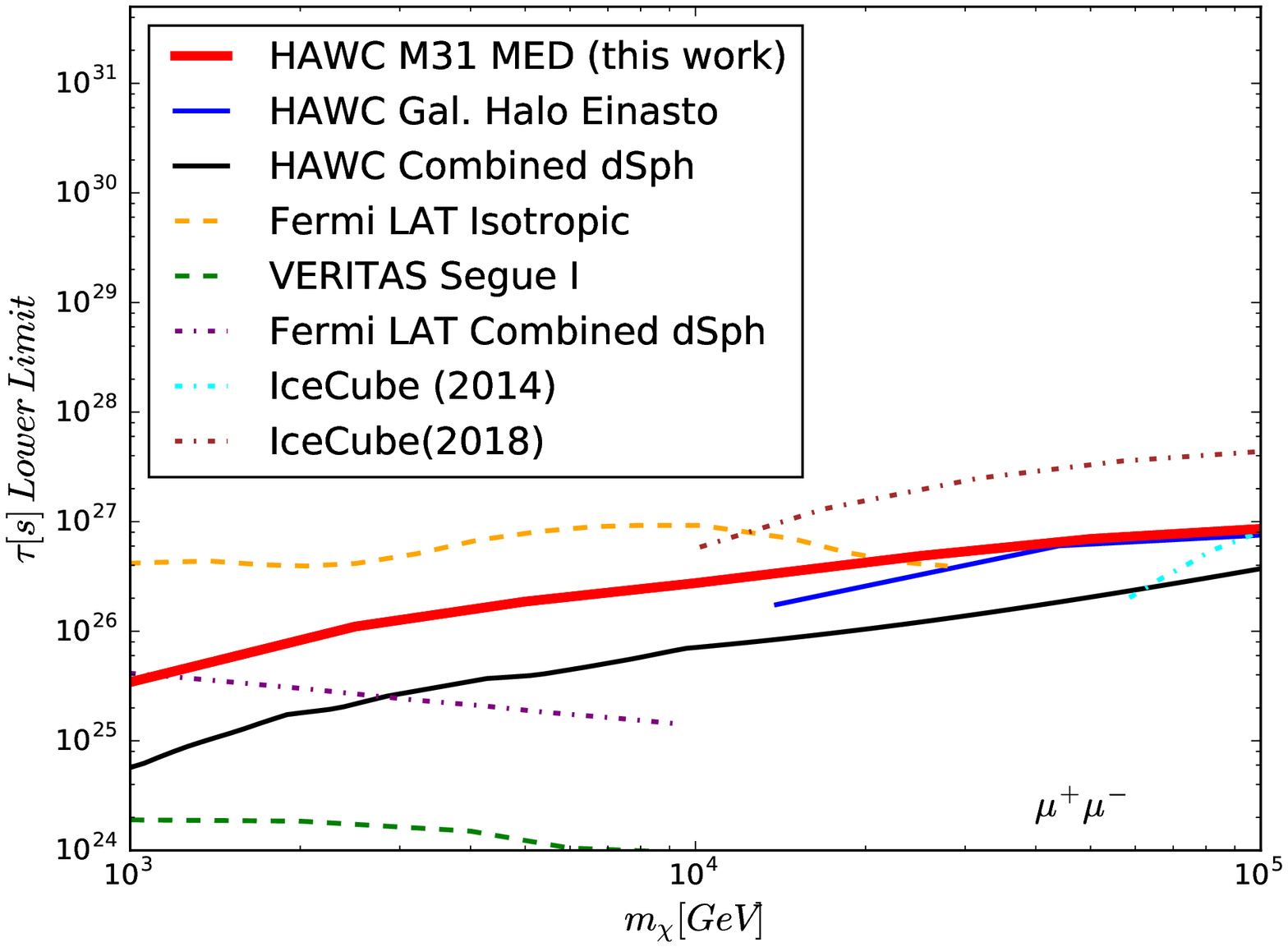}
    \includegraphics[width=0.48\textwidth]{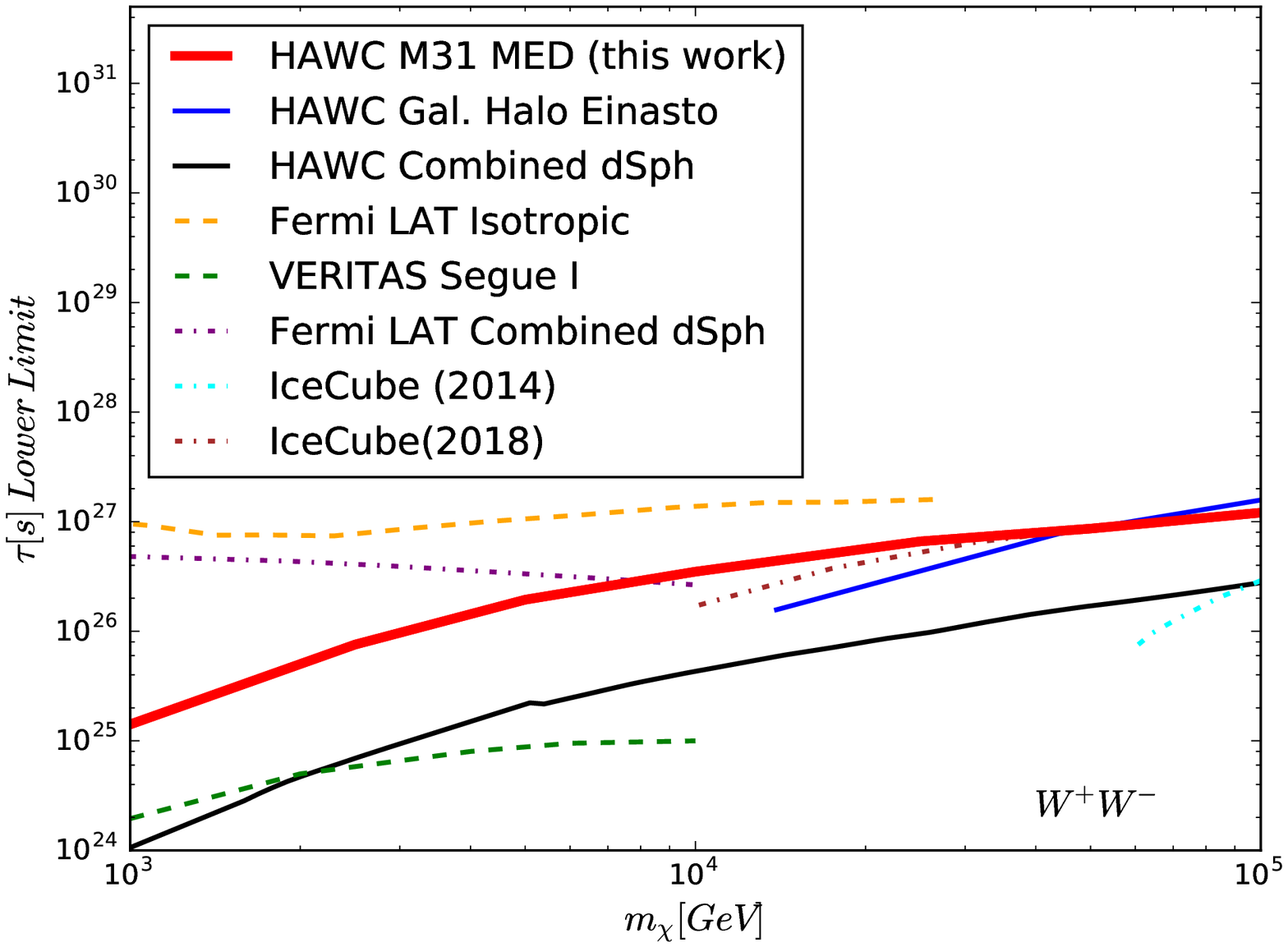}
    \caption{DM $\tau_{DM}$ 95\% confidence level lower limits for our benchmark DM halo model (MED,red) compared with recent DM searches with HAWC and other gamma-ray experiments when available for the $t\bar{t}$, $\tau^{+}\tau^{-}$, $\mu^{+}\mu^{-}$, and $W^{+}W^{-}$ channels. The limits obtained from the HAWC Galactic Halo search (specifically the northern \textit{Fermi} Bubble region)~\citep{harding} are shown in blue.  The HAWC limits obtained from a joint analysis of 15 dwarf spheroidal galaxies~\citep{Albert:2017vtb} are shown in black. The VERITAS limits obtained from 48 hours of observation on the dwarf spheroidal galaxy Segue I~\citep{Aliu:2012ga} are shown in green. The limits using \textit{Fermi} LAT data of 19 dwarf spheroidal galaxies~\citep{Baring:2015sza} are shown in purple. Limits from an analysis of the isotropic extragalactic background using \textit{Fermi} LAT data is shown in orange~\cite{Esmaili:2015xpa}. Decay limits from analyses of neutrinos by IceCube are shown in cyan~\cite{Esmaili:2014rma} and brown~\cite{Aartsen:2018mxl}. \label{fig:compDec2}}
\end{figure*}

The 95\% confidence level upper and lower limits for DM annihilation and decay producing gamma rays in the direction of M31 for the $t\bar{t}$, $\tau^{+}\tau^{-}$, $\mu^{+}\mu^{-}$, and $W^{+}W^{-}$ channels are shown in Figs~\ref{fig:allJAnn2} and \ref{fig:allJDec2}. Our limits for those channels compared to limits from other gamma-ray experiment are show in Figs~\ref{fig:compAnn2} and \ref{fig:compDec2}. All DM models' fit values are shown in Tabs~\ref{tab:limAnn} and \ref{tab:limDec}.

 \pagebreak
  
 \pagebreak
  
 \pagebreak
 
\begin{table*}[ht]
\begin{tabular}{ccccccc}
\hline \hline
DM mass [TeV] & DM Halo Model & $b\bar{b}$ & $t\bar{t}$ & $\tau^{+}\tau^{-}$ & $\mu^{+}\mu^{-}$ & $W^{+}W^{-}$ \\
 & & $(10^{-23}cm^{3}s^{-1})$ & $(10^{-23}cm^{3}s^{-1})$ & $(10^{-23}cm^{3}s^{-1})$ & $(10^{-23}cm^{3}s^{-1})$ & $(10^{-23}cm^{3}s^{-1})$ \\
\hline
1.0 & MIN & 132 & 180 & 17.2 & 47.5 & 122 \\
2.5 & MIN & 78.2 & 109 & 10.7 & 33.4 & 78.4  \\
5.0 & MIN & 53.1 & 76.5 & 10.4 & 35.0 & 53.0   \\
10.0 & MIN & 44.4 & 64.2 & 12.2 & 43.3 & 44.1  \\
25.0 & MIN & 44.2 & 64.2 & 18.5 & 65.6 & 44.5  \\
50.0 & MIN & 53.7 & 74.4 & 27.3 & 97.5 & 53.5\\
100.0 & MIN & 72.4 & 96.0 & 43.1 & 136 & 72.6 \\
1.0 & MED & 28.8 & 43.7 & 3.15 & 7.96 & 28.9  \\
2.5 & MED & 12.6 & 18.2 & 2.26 & 6.15 & 12.4   \\
5.0 & MED & 8.99 & 13.0 & 2.32 & 6.94 & 8.95  \\
10.0 & MED & 7.90 & 11.4 & 2.78 & 8.87 & 7.95 \\
25.0 & MED & 8.33 & 11.7 & 3.87 & 12.0 & 8.31 \\
50.0 & MED & 10.3 & 13.8 & 5.12 & 16.4 & 10.4  \\
100.0 & MED & 13.7 & 17.8 & 7.28 & 24.2 & 13.8 \\
1.0 & MAX & 6.14 & 9.25 & 0.571 & 1.52 & 6.12 \\
2.5 & MAX & 2.52 & 3.63 & 0.372 & 1.10 & 2.55 \\
5.0 & MAX & 1.72 & 2.50 & 0.367 & 1.17 & 1.75 \\
10.0 & MAX & 1.46 & 2.11 & 0.424 & 1.42 & 1.45 \\
25.0 & MAX & 1.48 & 2.16 & 0.632 & 2.22 & 1.43 \\
50.0 & MAX & 1.81 & 2.54 & 0.942 & 3.34 & 1.84 \\
100.0 & MAX & 2.47 & 3.30 & 1.45 & 5.03 & 2.48  \\
\end{tabular}
\caption{DM $\langle \sigma v\rangle$ 95\% confidence level upper limits for all DM models tested.
\label{tab:limAnn}}
\end{table*}

\begin{table*}[ht]
\begin{tabular}{ccccccc}
\hline \hline
DM mass [TeV] & DM Halo Model & $b\bar{b}$ & $t\bar{t}$ & $\tau^{+}\tau^{-}$ & $\mu^{+}\mu^{-}$ & $W^{+}W^{-}$ \\
 & & ($10^{26}s$) & ($10^{26}s$) & ($10^{26}s$) & ($10^{26}s$) & ($10^{26}s$) \\
\hline
1.0 & MIN & 0.0668 & 0.0520 & 0.755 & 0.285 & 0.115  \\
2.5 & MIN & 0.390 & 0.296 & 2.89 & 0.950 & 0.632   \\
5.0 & MIN & 1.12 & 0.800 & 5.56 & 1.67 & 1.69   \\
10.0 & MIN & 2.71 & 1.81 & 8.71 & 2.53 & 3.86  \\
25.0 & MIN & 6.18 & 4.45 & 14.9 & 4.74 & 7.71   \\
50.0 & MIN & 9.91 & 7.55 & 22.3 & 7.22 & 11.9 \\
100.0 & MIN & 14.8 & 11.6 & 30.2 & 9.35 & 14.9 \\
1.0 & MED & 0.0822 & 0.0638 & 0.903 & 0.343 & 0.141 \\
2.5 & MED & 0.471 & 0.357 & 3.21 & 1.10 & 0.756 \\
5.0 & MED & 1.31 & 0.942 & 5.71 & 1.87 & 1.94 \\
10.0 & MED & 3.02 & 2.06 & 8.28 & 2.75 & 3.50 \\
25.0 & MED & 6.24 & 4.67 & 12.5 & 4.87 & 6.61 \\
50.0 & MED & 9.14 & 7.31 & 16.1 & 6.96 & 8.64 \\
100.0 & MED & 12.3 & 10.2 & 19.0 & 8.60 & 12.1 \\ 
1.0 & MAX & 0.0521 & 0.0410 & 0.599 & 0.230 & 0.0891 \\
2.5 & MAX & 0.309 & 0.236 & 2.23 & 0.760 & 0.413 \\
5.0 & MAX & 0.883 & 0.632 & 4.13 & 1.31 & 1.22 \\
10.0 & MAX & 2.09 & 1.41 & 6.24 & 1.97 & 2.14 \\
25.0 & MAX & 4.52 & 3.28 & 9.68 & 3.35 & 4.53 \\
50.0 & MAX & 6.84 & 5.29 & 13.0 & 4.83 & 6.34 \\
100.0 & MAX & 9.49 & 7.65 & 16.0 & 6.17 & 8.68 \\
\end{tabular}
\caption{DM $\tau_{DM}$ 95\% confidence level lower limits for all DM models tested.
\label{tab:limDec}}
\end{table*}

%% file: M31-DM.bbl
 \newcommand{\noop}[1]{}
\begin{thebibliography}{69}%
\makeatletter
\providecommand \@ifxundefined [1]{%
 \@ifx{#1\undefined}
}%
\providecommand \@ifnum [1]{%
 \ifnum #1\expandafter \@firstoftwo
 \else \expandafter \@secondoftwo
 \fi
}%
\providecommand \@ifx [1]{%
 \ifx #1\expandafter \@firstoftwo
 \else \expandafter \@secondoftwo
 \fi
}%
\providecommand \natexlab [1]{#1}%
\providecommand \enquote  [1]{``#1''}%
\providecommand \bibnamefont  [1]{#1}%
\providecommand \bibfnamefont [1]{#1}%
\providecommand \citenamefont [1]{#1}%
\providecommand \href@noop [0]{\@secondoftwo}%
\providecommand \href [0]{\begingroup \@sanitize@url \@href}%
\providecommand \@href[1]{\@@startlink{#1}\@@href}%
\providecommand \@@href[1]{\endgroup#1\@@endlink}%
\providecommand \@sanitize@url [0]{\catcode `\\12\catcode `\$12\catcode
  `\&12\catcode `\#12\catcode `\^12\catcode `\_12\catcode `\%12\relax}%
\providecommand \@@startlink[1]{}%
\providecommand \@@endlink[0]{}%
\providecommand \url  [0]{\begingroup\@sanitize@url \@url }%
\providecommand \@url [1]{\endgroup\@href {#1}{\urlprefix }}%
\providecommand \urlprefix  [0]{URL }%
\providecommand \Eprint [0]{\href }%
\providecommand \doibase [0]{http://dx.doi.org/}%
\providecommand \selectlanguage [0]{\@gobble}%
\providecommand \bibinfo  [0]{\@secondoftwo}%
\providecommand \bibfield  [0]{\@secondoftwo}%
\providecommand \translation [1]{[#1]}%
\providecommand \BibitemOpen [0]{}%
\providecommand \bibitemStop [0]{}%
\providecommand \bibitemNoStop [0]{.\EOS\space}%
\providecommand \EOS [0]{\spacefactor3000\relax}%
\providecommand \BibitemShut  [1]{\csname bibitem#1\endcsname}%
\let\auto@bib@innerbib\@empty
%</preamble>
\bibitem [{\citenamefont {Ade}\ \emph {et~al.}(2014)\citenamefont {Ade} \emph
  {et~al.}}]{Ade:2013zuv}%
  \BibitemOpen
  \bibfield  {author} {\bibinfo {author} {\bibfnamefont {P.~A.~R.}\
  \bibnamefont {Ade}} \emph {et~al.} (\bibinfo {collaboration} {Planck}),\
  }\href {\doibase 10.1051/0004-6361/201321591} {\bibfield  {journal} {\bibinfo
   {journal} {Astron. Astrophys.}\ }\textbf {\bibinfo {volume} {571}},\
  \bibinfo {pages} {A16} (\bibinfo {year} {2014})},\ \Eprint
  {http://arxiv.org/abs/1303.5076} {arXiv:1303.5076 [astro-ph.CO]} \BibitemShut
  {NoStop}%
%%CITATION = ARXIV:1303.5076;%%
\bibitem [{\citenamefont {Clowe}\ \emph {et~al.}(2006)\citenamefont {Clowe},
  \citenamefont {Bradac}, \citenamefont {Gonzalez}, \citenamefont {Markevitch},
  \citenamefont {Randall}, \citenamefont {Jones},\ and\ \citenamefont
  {Zaritsky}}]{Clowe:2006eq}%
  \BibitemOpen
  \bibfield  {author} {\bibinfo {author} {\bibfnamefont {D.}~\bibnamefont
  {Clowe}}, \bibinfo {author} {\bibfnamefont {M.}~\bibnamefont {Bradac}},
  \bibinfo {author} {\bibfnamefont {A.~H.}\ \bibnamefont {Gonzalez}}, \bibinfo
  {author} {\bibfnamefont {M.}~\bibnamefont {Markevitch}}, \bibinfo {author}
  {\bibfnamefont {S.~W.}\ \bibnamefont {Randall}}, \bibinfo {author}
  {\bibfnamefont {C.}~\bibnamefont {Jones}}, \ and\ \bibinfo {author}
  {\bibfnamefont {D.}~\bibnamefont {Zaritsky}},\ }\href {\doibase
  10.1086/508162} {\bibfield  {journal} {\bibinfo  {journal} {Astrophys. J.}\
  }\textbf {\bibinfo {volume} {648}},\ \bibinfo {pages} {L109} (\bibinfo {year}
  {2006})},\ \Eprint {http://arxiv.org/abs/astro-ph/0608407}
  {arXiv:astro-ph/0608407 [astro-ph]} \BibitemShut {NoStop}%
%%CITATION = ASTRO-PH/0608407;%%
\bibitem [{\citenamefont {Sofue}\ and\ \citenamefont
  {Rubin}(2001)}]{Sofue:2000jx}%
  \BibitemOpen
  \bibfield  {author} {\bibinfo {author} {\bibfnamefont {Y.}~\bibnamefont
  {Sofue}}\ and\ \bibinfo {author} {\bibfnamefont {V.}~\bibnamefont {Rubin}},\
  }\href {\doibase 10.1146/annurev.astro.39.1.137} {\bibfield  {journal}
  {\bibinfo  {journal} {Ann. Rev. Astron. Astrophys.}\ }\textbf {\bibinfo
  {volume} {39}},\ \bibinfo {pages} {137} (\bibinfo {year} {2001})},\ \Eprint
  {http://arxiv.org/abs/astro-ph/0010594} {arXiv:astro-ph/0010594 [astro-ph]}
  \BibitemShut {NoStop}%
%%CITATION = ASTRO-PH/0010594;%%
\bibitem [{\citenamefont {Feng}(2010)}]{Feng:2010gw}%
  \BibitemOpen
  \bibfield  {author} {\bibinfo {author} {\bibfnamefont {J.~L.}\ \bibnamefont
  {Feng}},\ }\href {\doibase 10.1146/annurev-astro-082708-101659} {\bibfield
  {journal} {\bibinfo  {journal} {Ann. Rev. Astron. Astrophys.}\ }\textbf
  {\bibinfo {volume} {48}},\ \bibinfo {pages} {495} (\bibinfo {year} {2010})},\
  \Eprint {http://arxiv.org/abs/1003.0904} {arXiv:1003.0904 [astro-ph.CO]}
  \BibitemShut {NoStop}%
%%CITATION = ARXIV:1003.0904;%%
\bibitem [{\citenamefont {Baer}\ \emph {et~al.}(2015)\citenamefont {Baer},
  \citenamefont {Choi}, \citenamefont {Kim},\ and\ \citenamefont
  {Roszkowski}}]{Baer:2014eja}%
  \BibitemOpen
  \bibfield  {author} {\bibinfo {author} {\bibfnamefont {H.}~\bibnamefont
  {Baer}}, \bibinfo {author} {\bibfnamefont {K.-Y.}\ \bibnamefont {Choi}},
  \bibinfo {author} {\bibfnamefont {J.~E.}\ \bibnamefont {Kim}}, \ and\
  \bibinfo {author} {\bibfnamefont {L.}~\bibnamefont {Roszkowski}},\ }\href
  {\doibase 10.1016/j.physrep.2014.10.002} {\bibfield  {journal} {\bibinfo
  {journal} {Phys. Rept.}\ }\textbf {\bibinfo {volume} {555}},\ \bibinfo
  {pages} {1} (\bibinfo {year} {2015})},\ \Eprint
  {http://arxiv.org/abs/1407.0017} {arXiv:1407.0017 [hep-ph]} \BibitemShut
  {NoStop}%
%%CITATION = ARXIV:1407.0017;%%
\bibitem [{\citenamefont {Carr}\ \emph {et~al.}(2016)\citenamefont {Carr},
  \citenamefont {Kuhnel},\ and\ \citenamefont {Sandstad}}]{Carr:2016drx}%
  \BibitemOpen
  \bibfield  {author} {\bibinfo {author} {\bibfnamefont {B.}~\bibnamefont
  {Carr}}, \bibinfo {author} {\bibfnamefont {F.}~\bibnamefont {Kuhnel}}, \ and\
  \bibinfo {author} {\bibfnamefont {M.}~\bibnamefont {Sandstad}},\ }\href
  {\doibase 10.1103/PhysRevD.94.083504} {\bibfield  {journal} {\bibinfo
  {journal} {Phys. Rev.}\ }\textbf {\bibinfo {volume} {D94}},\ \bibinfo {pages}
  {083504} (\bibinfo {year} {2016})},\ \Eprint
  {http://arxiv.org/abs/1607.06077} {arXiv:1607.06077 [astro-ph.CO]}
  \BibitemShut {NoStop}%
%%CITATION = ARXIV:1607.06077;%%
\bibitem [{\citenamefont {Bird}\ \emph {et~al.}(2016)\citenamefont {Bird},
  \citenamefont {Cholis}, \citenamefont {Muñoz}, \citenamefont {Ali-Haïmoud},
  \citenamefont {Kamionkowski}, \citenamefont {Kovetz}, \citenamefont
  {Raccanelli},\ and\ \citenamefont {Riess}}]{Bird:2016dcv}%
  \BibitemOpen
  \bibfield  {author} {\bibinfo {author} {\bibfnamefont {S.}~\bibnamefont
  {Bird}}, \bibinfo {author} {\bibfnamefont {I.}~\bibnamefont {Cholis}},
  \bibinfo {author} {\bibfnamefont {J.~B.}\ \bibnamefont {Muñoz}}, \bibinfo
  {author} {\bibfnamefont {Y.}~\bibnamefont {Ali-Haïmoud}}, \bibinfo {author}
  {\bibfnamefont {M.}~\bibnamefont {Kamionkowski}}, \bibinfo {author}
  {\bibfnamefont {E.~D.}\ \bibnamefont {Kovetz}}, \bibinfo {author}
  {\bibfnamefont {A.}~\bibnamefont {Raccanelli}}, \ and\ \bibinfo {author}
  {\bibfnamefont {A.~G.}\ \bibnamefont {Riess}},\ }\href {\doibase
  10.1103/PhysRevLett.116.201301} {\bibfield  {journal} {\bibinfo  {journal}
  {Phys. Rev. Lett.}\ }\textbf {\bibinfo {volume} {116}},\ \bibinfo {pages}
  {201301} (\bibinfo {year} {2016})},\ \Eprint
  {http://arxiv.org/abs/1603.00464} {arXiv:1603.00464 [astro-ph.CO]}
  \BibitemShut {NoStop}%
%%CITATION = ARXIV:1603.00464;%%
\bibitem [{\citenamefont {Peccei}\ and\ \citenamefont
  {Quinn}(1977)}]{Peccei:1977hh}%
  \BibitemOpen
  \bibfield  {author} {\bibinfo {author} {\bibfnamefont {R.~D.}\ \bibnamefont
  {Peccei}}\ and\ \bibinfo {author} {\bibfnamefont {H.~R.}\ \bibnamefont
  {Quinn}},\ }\href {\doibase 10.1103/PhysRevLett.38.1440} {\bibfield
  {journal} {\bibinfo  {journal} {Phys. Rev. Lett.}\ }\textbf {\bibinfo
  {volume} {38}},\ \bibinfo {pages} {1440} (\bibinfo {year}
  {1977})}\BibitemShut {NoStop}%
%%CITATION = PRLTA,38,1440;%%
\bibitem [{\citenamefont {Arias}\ \emph {et~al.}(2012)\citenamefont {Arias},
  \citenamefont {Cadamuro}, \citenamefont {Goodsell}, \citenamefont {Jaeckel},
  \citenamefont {Redondo},\ and\ \citenamefont {Ringwald}}]{Arias:2012az}%
  \BibitemOpen
  \bibfield  {author} {\bibinfo {author} {\bibfnamefont {P.}~\bibnamefont
  {Arias}}, \bibinfo {author} {\bibfnamefont {D.}~\bibnamefont {Cadamuro}},
  \bibinfo {author} {\bibfnamefont {M.}~\bibnamefont {Goodsell}}, \bibinfo
  {author} {\bibfnamefont {J.}~\bibnamefont {Jaeckel}}, \bibinfo {author}
  {\bibfnamefont {J.}~\bibnamefont {Redondo}}, \ and\ \bibinfo {author}
  {\bibfnamefont {A.}~\bibnamefont {Ringwald}},\ }\href {\doibase
  10.1088/1475-7516/2012/06/013} {\bibfield  {journal} {\bibinfo  {journal}
  {JCAP}\ }\textbf {\bibinfo {volume} {1206}},\ \bibinfo {pages} {013}
  (\bibinfo {year} {2012})},\ \Eprint {http://arxiv.org/abs/1201.5902}
  {arXiv:1201.5902 [hep-ph]} \BibitemShut {NoStop}%
%%CITATION = ARXIV:1201.5902;%%
\bibitem [{\citenamefont {Ackermann}\ \emph {et~al.}(2015)\citenamefont
  {Ackermann} \emph {et~al.}}]{Ackermann:2015zua}%
  \BibitemOpen
  \bibfield  {author} {\bibinfo {author} {\bibfnamefont {M.}~\bibnamefont
  {Ackermann}} \emph {et~al.} (\bibinfo {collaboration} {{Fermi-LAT
  Collaboration}}),\ }\href {\doibase 10.1103/PhysRevLett.115.231301}
  {\bibfield  {journal} {\bibinfo  {journal} {\prl}\ }\textbf {\bibinfo
  {volume} {115}},\ \bibinfo {pages} {231301} (\bibinfo {year} {2015})},\
  \Eprint {http://arxiv.org/abs/1503.02641} {arXiv:1503.02641 [astro-ph.HE]}
  \BibitemShut {NoStop}%
%%CITATION = ARXIV:1503.02641;%%
\bibitem [{\citenamefont {Aaboud}\ \emph
  {et~al.}(2017{\natexlab{a}})\citenamefont {Aaboud} \emph
  {et~al.}}]{Aaboud:2017dor}%
  \BibitemOpen
  \bibfield  {author} {\bibinfo {author} {\bibfnamefont {M.}~\bibnamefont
  {Aaboud}} \emph {et~al.} (\bibinfo {collaboration} {ATLAS}),\ }\href
  {\doibase 10.1140/epjc/s10052-017-4965-8} {\bibfield  {journal} {\bibinfo
  {journal} {Eur. Phys. J.}\ }\textbf {\bibinfo {volume} {C77}},\ \bibinfo
  {pages} {393} (\bibinfo {year} {2017}{\natexlab{a}})},\ \Eprint
  {http://arxiv.org/abs/1704.03848} {arXiv:1704.03848 [hep-ex]} \BibitemShut
  {NoStop}%
%%CITATION = ARXIV:1704.03848;%%
\bibitem [{\citenamefont {Aaboud}\ \emph
  {et~al.}(2017{\natexlab{b}})\citenamefont {Aaboud} \emph
  {et~al.}}]{Aaboud:2016obm}%
  \BibitemOpen
  \bibfield  {author} {\bibinfo {author} {\bibfnamefont {M.}~\bibnamefont
  {Aaboud}} \emph {et~al.} (\bibinfo {collaboration} {ATLAS}),\ }\href
  {\doibase 10.1016/j.physletb.2016.11.035} {\bibfield  {journal} {\bibinfo
  {journal} {Phys. Lett.}\ }\textbf {\bibinfo {volume} {B765}},\ \bibinfo
  {pages} {11} (\bibinfo {year} {2017}{\natexlab{b}})},\ \Eprint
  {http://arxiv.org/abs/1609.04572} {arXiv:1609.04572 [hep-ex]} \BibitemShut
  {NoStop}%
%%CITATION = ARXIV:1609.04572;%%
\bibitem [{\citenamefont {Sirunyan}\ \emph {et~al.}(2017)\citenamefont
  {Sirunyan} \emph {et~al.}}]{Sirunyan:2017onm}%
  \BibitemOpen
  \bibfield  {author} {\bibinfo {author} {\bibfnamefont {A.~M.}\ \bibnamefont
  {Sirunyan}} \emph {et~al.} (\bibinfo {collaboration} {CMS}),\ }\href
  {\doibase 10.1007/JHEP03(2017)061} {\bibfield  {journal} {\bibinfo  {journal}
  {JHEP}\ }\textbf {\bibinfo {volume} {03}},\ \bibinfo {pages} {061} (\bibinfo
  {year} {2017})},\ \Eprint {http://arxiv.org/abs/1701.02042} {arXiv:1701.02042
  [hep-ex]} \BibitemShut {NoStop}%
%%CITATION = ARXIV:1701.02042;%%
\bibitem [{\citenamefont {Khachatryan}\ \emph {et~al.}(2016)\citenamefont
  {Khachatryan} \emph {et~al.}}]{Khachatryan:2016mdm}%
  \BibitemOpen
  \bibfield  {author} {\bibinfo {author} {\bibfnamefont {V.}~\bibnamefont
  {Khachatryan}} \emph {et~al.} (\bibinfo {collaboration} {CMS}),\ }\href
  {\doibase 10.1007/JHEP12(2016)083} {\bibfield  {journal} {\bibinfo  {journal}
  {JHEP}\ }\textbf {\bibinfo {volume} {12}},\ \bibinfo {pages} {083} (\bibinfo
  {year} {2016})},\ \Eprint {http://arxiv.org/abs/1607.05764} {arXiv:1607.05764
  [hep-ex]} \BibitemShut {NoStop}%
%%CITATION = ARXIV:1607.05764;%%
\bibitem [{\citenamefont {Akerib}\ \emph {et~al.}(2017)\citenamefont {Akerib}
  \emph {et~al.}}]{Akerib:2017kat}%
  \BibitemOpen
  \bibfield  {author} {\bibinfo {author} {\bibfnamefont {D.~S.}\ \bibnamefont
  {Akerib}} \emph {et~al.} (\bibinfo {collaboration} {LUX}),\ }\href {\doibase
  10.1103/PhysRevLett.118.251302} {\bibfield  {journal} {\bibinfo  {journal}
  {Phys. Rev. Lett.}\ }\textbf {\bibinfo {volume} {118}},\ \bibinfo {pages}
  {251302} (\bibinfo {year} {2017})},\ \Eprint
  {http://arxiv.org/abs/1705.03380} {arXiv:1705.03380 [astro-ph.CO]}
  \BibitemShut {NoStop}%
%%CITATION = ARXIV:1705.03380;%%
\bibitem [{\citenamefont {Aprile}\ \emph {et~al.}(2016)\citenamefont {Aprile}
  \emph {et~al.}}]{Aprile:2016swn}%
  \BibitemOpen
  \bibfield  {author} {\bibinfo {author} {\bibfnamefont {E.}~\bibnamefont
  {Aprile}} \emph {et~al.} (\bibinfo {collaboration} {XENON100}),\ }\href
  {\doibase 10.1103/PhysRevD.94.122001} {\bibfield  {journal} {\bibinfo
  {journal} {Phys. Rev.}\ }\textbf {\bibinfo {volume} {D94}},\ \bibinfo {pages}
  {122001} (\bibinfo {year} {2016})},\ \Eprint
  {http://arxiv.org/abs/1609.06154} {arXiv:1609.06154 [astro-ph.CO]}
  \BibitemShut {NoStop}%
%%CITATION = ARXIV:1609.06154;%%
\bibitem [{\citenamefont {Blanco}\ \emph {et~al.}(2017)\citenamefont {Blanco},
  \citenamefont {Harding},\ and\ \citenamefont {Hooper}}]{Blanco:2017sbc}%
  \BibitemOpen
  \bibfield  {author} {\bibinfo {author} {\bibfnamefont {C.}~\bibnamefont
  {Blanco}}, \bibinfo {author} {\bibfnamefont {J.~P.}\ \bibnamefont {Harding}},
  \ and\ \bibinfo {author} {\bibfnamefont {D.}~\bibnamefont {Hooper}},\
  }\href@noop {} {\  (\bibinfo {year} {2017})},\ \Eprint
  {http://arxiv.org/abs/1712.02805} {arXiv:1712.02805 [hep-ph]} \BibitemShut
  {NoStop}%
%%CITATION = ARXIV:1712.02805;%%
\bibitem [{\citenamefont {Garcia-Cely}\ \emph {et~al.}(2015)\citenamefont
  {Garcia-Cely}, \citenamefont {Ibarra}, \citenamefont {Lamperstorfer},\ and\
  \citenamefont {Tytgat}}]{Garcia-Cely:2015dda}%
  \BibitemOpen
  \bibfield  {author} {\bibinfo {author} {\bibfnamefont {C.}~\bibnamefont
  {Garcia-Cely}}, \bibinfo {author} {\bibfnamefont {A.}~\bibnamefont {Ibarra}},
  \bibinfo {author} {\bibfnamefont {A.~S.}\ \bibnamefont {Lamperstorfer}}, \
  and\ \bibinfo {author} {\bibfnamefont {M.~H.~G.}\ \bibnamefont {Tytgat}},\
  }\href {\doibase 10.1088/1475-7516/2015/10/058} {\bibfield  {journal}
  {\bibinfo  {journal} {JCAP}\ }\textbf {\bibinfo {volume} {1510}},\ \bibinfo
  {pages} {058} (\bibinfo {year} {2015})},\ \Eprint
  {http://arxiv.org/abs/1507.05536} {arXiv:1507.05536 [hep-ph]} \BibitemShut
  {NoStop}%
%%CITATION = ARXIV:1507.05536;%%
\bibitem [{\citenamefont {Cholis}\ \emph {et~al.}(2009)\citenamefont {Cholis},
  \citenamefont {Finkbeiner}, \citenamefont {Goodenough},\ and\ \citenamefont
  {Weiner}}]{Cholis:2008qq}%
  \BibitemOpen
  \bibfield  {author} {\bibinfo {author} {\bibfnamefont {I.}~\bibnamefont
  {Cholis}}, \bibinfo {author} {\bibfnamefont {D.~P.}\ \bibnamefont
  {Finkbeiner}}, \bibinfo {author} {\bibfnamefont {L.}~\bibnamefont
  {Goodenough}}, \ and\ \bibinfo {author} {\bibfnamefont {N.}~\bibnamefont
  {Weiner}},\ }\href {\doibase 10.1088/1475-7516/2009/12/007} {\bibfield
  {journal} {\bibinfo  {journal} {JCAP}\ }\textbf {\bibinfo {volume} {0912}},\
  \bibinfo {pages} {007} (\bibinfo {year} {2009})},\ \Eprint
  {http://arxiv.org/abs/0810.5344} {arXiv:0810.5344 [astro-ph]} \BibitemShut
  {NoStop}%
%%CITATION = ARXIV:0810.5344;%%
\bibitem [{\citenamefont {Tamm}\ \emph {et~al.}(2012)\citenamefont {Tamm} \emph
  {et~al.}}]{tamm2012stellar}%
  \BibitemOpen
  \bibfield  {author} {\bibinfo {author} {\bibfnamefont {A.}~\bibnamefont
  {Tamm}} \emph {et~al.},\ }\href@noop {} {\bibfield  {journal} {\bibinfo
  {journal} {Astronomy \& Astrophysics}\ }\textbf {\bibinfo {volume} {546}},\
  \bibinfo {pages} {A4} (\bibinfo {year} {2012})}\BibitemShut {NoStop}%
\bibitem [{\citenamefont {Tollerud}\ \emph {et~al.}(2012)\citenamefont
  {Tollerud} \emph {et~al.}}]{Tollerud:2011mi}%
  \BibitemOpen
  \bibfield  {author} {\bibinfo {author} {\bibfnamefont {E.~J.}\ \bibnamefont
  {Tollerud}} \emph {et~al.},\ }\href {\doibase 10.1088/0004-637X/752/1/45}
  {\bibfield  {journal} {\bibinfo  {journal} {Astrophys. J.}\ }\textbf
  {\bibinfo {volume} {752}},\ \bibinfo {pages} {45} (\bibinfo {year} {2012})},\
  \Eprint {http://arxiv.org/abs/1112.1067} {arXiv:1112.1067 [astro-ph.CO]}
  \BibitemShut {NoStop}%
%%CITATION = ARXIV:1112.1067;%%
\bibitem [{\citenamefont {Ackermann}\ \emph {et~al.}(2017)\citenamefont
  {Ackermann} \emph {et~al.}}]{M31Fermi}%
  \BibitemOpen
  \bibfield  {author} {\bibinfo {author} {\bibfnamefont {M.}~\bibnamefont
  {Ackermann}} \emph {et~al.} (\bibinfo {collaboration} {Fermi-LAT}),\ }\href
  {\doibase 10.3847/1538-4357/aa5c3d} {\bibfield  {journal} {\bibinfo
  {journal} {Astrophys. J.}\ }\textbf {\bibinfo {volume} {836}},\ \bibinfo
  {pages} {208} (\bibinfo {year} {2017})},\ \Eprint
  {http://arxiv.org/abs/1702.08602} {arXiv:1702.08602 [astro-ph.HE]}
  \BibitemShut {NoStop}%
%%CITATION = ARXIV:1702.08602;%%
\bibitem [{\citenamefont {McDaniel}\ \emph {et~al.}(2018)\citenamefont
  {McDaniel}, \citenamefont {Jeltema},\ and\ \citenamefont
  {Profumo}}]{McDaniel:2018vam}%
  \BibitemOpen
  \bibfield  {author} {\bibinfo {author} {\bibfnamefont {A.}~\bibnamefont
  {McDaniel}}, \bibinfo {author} {\bibfnamefont {T.}~\bibnamefont {Jeltema}}, \
  and\ \bibinfo {author} {\bibfnamefont {S.}~\bibnamefont {Profumo}},\
  }\href@noop {} {\  (\bibinfo {year} {2018})},\ \Eprint
  {http://arxiv.org/abs/1802.05258} {arXiv:1802.05258 [astro-ph.HE]}
  \BibitemShut {NoStop}%
%%CITATION = ARXIV:1802.05258;%%
\bibitem [{\citenamefont {Vianello}\ \emph {et~al.}(2015)\citenamefont
  {Vianello}, \citenamefont {Lauer}, \citenamefont {Younk}, \citenamefont
  {Tibaldo}, \citenamefont {Burgess}, \citenamefont {Ayala}, \citenamefont
  {Harding}, \citenamefont {Hui}, \citenamefont {Omodei},\ and\ \citenamefont
  {Zhou}}]{Vianello:2015wwa}%
  \BibitemOpen
  \bibfield  {author} {\bibinfo {author} {\bibfnamefont {G.}~\bibnamefont
  {Vianello}}, \bibinfo {author} {\bibfnamefont {R.~J.}\ \bibnamefont {Lauer}},
  \bibinfo {author} {\bibfnamefont {P.}~\bibnamefont {Younk}}, \bibinfo
  {author} {\bibfnamefont {L.}~\bibnamefont {Tibaldo}}, \bibinfo {author}
  {\bibfnamefont {J.~M.}\ \bibnamefont {Burgess}}, \bibinfo {author}
  {\bibfnamefont {H.}~\bibnamefont {Ayala}}, \bibinfo {author} {\bibfnamefont
  {P.}~\bibnamefont {Harding}}, \bibinfo {author} {\bibfnamefont
  {M.}~\bibnamefont {Hui}}, \bibinfo {author} {\bibfnamefont {N.}~\bibnamefont
  {Omodei}}, \ and\ \bibinfo {author} {\bibfnamefont {H.}~\bibnamefont {Zhou}}\
  }(\bibinfo {year} {2015})\ \Eprint {http://arxiv.org/abs/1507.08343}
  {arXiv:1507.08343 [astro-ph.HE]} \BibitemShut {NoStop}%
%%CITATION = ARXIV:1507.08343;%%
\bibitem [{\citenamefont {Jeltema}\ and\ \citenamefont
  {Profumo}(2008)}]{Jeltema:2008hf}%
  \BibitemOpen
  \bibfield  {author} {\bibinfo {author} {\bibfnamefont {T.~E.}\ \bibnamefont
  {Jeltema}}\ and\ \bibinfo {author} {\bibfnamefont {S.}~\bibnamefont
  {Profumo}},\ }\href {\doibase 10.1088/1475-7516/2008/11/003} {\bibfield
  {journal} {\bibinfo  {journal} {JCAP}\ }\textbf {\bibinfo {volume} {0811}},\
  \bibinfo {pages} {003} (\bibinfo {year} {2008})},\ \Eprint
  {http://arxiv.org/abs/0808.2641} {arXiv:0808.2641 [astro-ph]} \BibitemShut
  {NoStop}%
%%CITATION = ARXIV:0808.2641;%%
\bibitem [{\citenamefont {Albert}\ \emph {et~al.}(2018)\citenamefont {Albert}
  \emph {et~al.}}]{Albert:2017vtb}%
  \BibitemOpen
  \bibfield  {author} {\bibinfo {author} {\bibfnamefont {A.}~\bibnamefont
  {Albert}} \emph {et~al.} (\bibinfo {collaboration} {HAWC}),\ }\href {\doibase
  10.3847/1538-4357/aaa6d8} {\bibfield  {journal} {\bibinfo  {journal}
  {Astrophys. J.}\ }\textbf {\bibinfo {volume} {853}},\ \bibinfo {pages} {154}
  (\bibinfo {year} {2018})},\ \Eprint {http://arxiv.org/abs/1706.01277}
  {arXiv:1706.01277 [astro-ph.HE]} \BibitemShut {NoStop}%
%%CITATION = ARXIV:1706.01277;%%
\bibitem [{\citenamefont {Sjostrand}\ \emph {et~al.}(2006)\citenamefont
  {Sjostrand}, \citenamefont {Mrenna},\ and\ \citenamefont
  {Skands}}]{Sjostrand:2006za}%
  \BibitemOpen
  \bibfield  {author} {\bibinfo {author} {\bibfnamefont {T.}~\bibnamefont
  {Sjostrand}}, \bibinfo {author} {\bibfnamefont {S.}~\bibnamefont {Mrenna}}, \
  and\ \bibinfo {author} {\bibfnamefont {P.~Z.}\ \bibnamefont {Skands}},\
  }\href {\doibase 10.1088/1126-6708/2006/05/026} {\bibfield  {journal}
  {\bibinfo  {journal} {JHEP}\ }\textbf {\bibinfo {volume} {05}},\ \bibinfo
  {pages} {026} (\bibinfo {year} {2006})},\ \Eprint
  {http://arxiv.org/abs/hep-ph/0603175} {arXiv:hep-ph/0603175 [hep-ph]}
  \BibitemShut {NoStop}%
%%CITATION = HEP-PH/0603175;%%
\bibitem [{\citenamefont {Sjöstrand}\ \emph {et~al.}(2015)\citenamefont
  {Sjöstrand}, \citenamefont {Ask}, \citenamefont {Christiansen},
  \citenamefont {Corke}, \citenamefont {Desai}, \citenamefont {Ilten},
  \citenamefont {Mrenna}, \citenamefont {Prestel}, \citenamefont {Rasmussen},\
  and\ \citenamefont {Skands}}]{Sjostrand:2014zea}%
  \BibitemOpen
  \bibfield  {author} {\bibinfo {author} {\bibfnamefont {T.}~\bibnamefont
  {Sjöstrand}}, \bibinfo {author} {\bibfnamefont {S.}~\bibnamefont {Ask}},
  \bibinfo {author} {\bibfnamefont {J.~R.}\ \bibnamefont {Christiansen}},
  \bibinfo {author} {\bibfnamefont {R.}~\bibnamefont {Corke}}, \bibinfo
  {author} {\bibfnamefont {N.}~\bibnamefont {Desai}}, \bibinfo {author}
  {\bibfnamefont {P.}~\bibnamefont {Ilten}}, \bibinfo {author} {\bibfnamefont
  {S.}~\bibnamefont {Mrenna}}, \bibinfo {author} {\bibfnamefont
  {S.}~\bibnamefont {Prestel}}, \bibinfo {author} {\bibfnamefont {C.~O.}\
  \bibnamefont {Rasmussen}}, \ and\ \bibinfo {author} {\bibfnamefont {P.~Z.}\
  \bibnamefont {Skands}},\ }\href {\doibase 10.1016/j.cpc.2015.01.024}
  {\bibfield  {journal} {\bibinfo  {journal} {Comput. Phys. Commun.}\ }\textbf
  {\bibinfo {volume} {191}},\ \bibinfo {pages} {159} (\bibinfo {year}
  {2015})},\ \Eprint {http://arxiv.org/abs/1410.3012} {arXiv:1410.3012
  [hep-ph]} \BibitemShut {NoStop}%
%%CITATION = ARXIV:1410.3012;%%
\bibitem [{\citenamefont {Esmaili}\ and\ \citenamefont
  {Serpico}(2015)}]{Esmaili:2015xpa}%
  \BibitemOpen
  \bibfield  {author} {\bibinfo {author} {\bibfnamefont {A.}~\bibnamefont
  {Esmaili}}\ and\ \bibinfo {author} {\bibfnamefont {P.~D.}\ \bibnamefont
  {Serpico}},\ }\href {\doibase 10.1088/1475-7516/2015/10/014} {\bibfield
  {journal} {\bibinfo  {journal} {JCAP}\ }\textbf {\bibinfo {volume} {1510}},\
  \bibinfo {pages} {014} (\bibinfo {year} {2015})},\ \Eprint
  {http://arxiv.org/abs/1505.06486} {arXiv:1505.06486 [hep-ph]} \BibitemShut
  {NoStop}%
%%CITATION = ARXIV:1505.06486;%%
\bibitem [{\citenamefont {Bonnivard}\ \emph {et~al.}(2016)\citenamefont
  {Bonnivard} \emph {et~al.}}]{bonnivard2016clumpy}%
  \BibitemOpen
  \bibfield  {author} {\bibinfo {author} {\bibfnamefont {V.}~\bibnamefont
  {Bonnivard}} \emph {et~al.},\ }\href@noop {} {\bibfield  {journal} {\bibinfo
  {journal} {Computer physics communications}\ }\textbf {\bibinfo {volume}
  {200}},\ \bibinfo {pages} {336} (\bibinfo {year} {2016})}\BibitemShut
  {NoStop}%
\bibitem [{\citenamefont {Charbonnier}\ \emph {et~al.}(2012)\citenamefont
  {Charbonnier}, \citenamefont {Combet},\ and\ \citenamefont
  {Maurin}}]{charbonnier2012clumpy}%
  \BibitemOpen
  \bibfield  {author} {\bibinfo {author} {\bibfnamefont {A.}~\bibnamefont
  {Charbonnier}}, \bibinfo {author} {\bibfnamefont {C.}~\bibnamefont {Combet}},
  \ and\ \bibinfo {author} {\bibfnamefont {D.}~\bibnamefont {Maurin}},\
  }\href@noop {} {\bibfield  {journal} {\bibinfo  {journal} {Computer Physics
  Communications}\ }\textbf {\bibinfo {volume} {183}},\ \bibinfo {pages} {656}
  (\bibinfo {year} {2012})}\BibitemShut {NoStop}%
\bibitem [{\citenamefont {Springel}\ \emph {et~al.}(2008)\citenamefont
  {Springel} \emph {et~al.}}]{springel2008aquarius}%
  \BibitemOpen
  \bibfield  {author} {\bibinfo {author} {\bibfnamefont {V.}~\bibnamefont
  {Springel}} \emph {et~al.},\ }\href@noop {} {\bibfield  {journal} {\bibinfo
  {journal} {Monthly Notices of the Royal Astronomical Society}\ }\textbf
  {\bibinfo {volume} {391}},\ \bibinfo {pages} {1685} (\bibinfo {year}
  {2008})}\BibitemShut {NoStop}%
\bibitem [{\citenamefont {Kuhlen}\ \emph
  {et~al.}(2008{\natexlab{a}})\citenamefont {Kuhlen}, \citenamefont {Diemand},
  \citenamefont {Madau},\ and\ \citenamefont {Zemp}}]{kuhlen2008via}%
  \BibitemOpen
  \bibfield  {author} {\bibinfo {author} {\bibfnamefont {M.}~\bibnamefont
  {Kuhlen}}, \bibinfo {author} {\bibfnamefont {J.}~\bibnamefont {Diemand}},
  \bibinfo {author} {\bibfnamefont {P.}~\bibnamefont {Madau}}, \ and\ \bibinfo
  {author} {\bibfnamefont {M.}~\bibnamefont {Zemp}},\ }in\ \href@noop {} {\emph
  {\bibinfo {booktitle} {Journal of Physics: Conference Series}}},\ Vol.\
  \bibinfo {volume} {125}\ (\bibinfo {organization} {IOP Publishing},\ \bibinfo
  {year} {2008})\ p.\ \bibinfo {pages} {012008}\BibitemShut {NoStop}%
\bibitem [{\citenamefont {Griffen}\ \emph {et~al.}(2016)\citenamefont {Griffen}
  \emph {et~al.}}]{griffen2016caterpillar}%
  \BibitemOpen
  \bibfield  {author} {\bibinfo {author} {\bibfnamefont {B.~F.}\ \bibnamefont
  {Griffen}} \emph {et~al.},\ }\href@noop {} {\bibfield  {journal} {\bibinfo
  {journal} {The Astrophysical Journal}\ }\textbf {\bibinfo {volume} {818}},\
  \bibinfo {pages} {10} (\bibinfo {year} {2016})}\BibitemShut {NoStop}%
\bibitem [{\citenamefont {Kuhlen}\ \emph
  {et~al.}(2008{\natexlab{b}})\citenamefont {Kuhlen}, \citenamefont {Diemand},\
  and\ \citenamefont {Madau}}]{kuhlen2008dark}%
  \BibitemOpen
  \bibfield  {author} {\bibinfo {author} {\bibfnamefont {M.}~\bibnamefont
  {Kuhlen}}, \bibinfo {author} {\bibfnamefont {J.}~\bibnamefont {Diemand}}, \
  and\ \bibinfo {author} {\bibfnamefont {P.}~\bibnamefont {Madau}},\
  }\href@noop {} {\bibfield  {journal} {\bibinfo  {journal} {The Astrophysical
  Journal}\ }\textbf {\bibinfo {volume} {686}},\ \bibinfo {pages} {262}
  (\bibinfo {year} {2008}{\natexlab{b}})}\BibitemShut {NoStop}%
\bibitem [{\citenamefont {Sanchez-Conde}\ and\ \citenamefont
  {Prada}(2014)}]{Sanchez-Conde:2013yxa}%
  \BibitemOpen
  \bibfield  {author} {\bibinfo {author} {\bibfnamefont {M.}~\bibnamefont
  {Sanchez-Conde}}\ and\ \bibinfo {author} {\bibfnamefont {F.}~\bibnamefont
  {Prada}},\ }\href {\doibase 10.1093/mnras/stu1014} {\bibfield  {journal}
  {\bibinfo  {journal} {Mon. Not. Roy. Astron. Soc.}\ }\textbf {\bibinfo
  {volume} {442}},\ \bibinfo {pages} {2271} (\bibinfo {year} {2014})},\ \Eprint
  {http://arxiv.org/abs/1312.1729} {arXiv:1312.1729 [astro-ph.CO]} \BibitemShut
  {NoStop}%
%%CITATION = ARXIV:1312.1729;%%
\bibitem [{\citenamefont {Bullock}\ \emph {et~al.}(2001)\citenamefont {Bullock}
  \emph {et~al.}}]{bullock2001profiles}%
  \BibitemOpen
  \bibfield  {author} {\bibinfo {author} {\bibfnamefont {J.~S.}\ \bibnamefont
  {Bullock}} \emph {et~al.},\ }\href@noop {} {\bibfield  {journal} {\bibinfo
  {journal} {Monthly Notices of the Royal Astronomical Society}\ }\textbf
  {\bibinfo {volume} {321}},\ \bibinfo {pages} {559} (\bibinfo {year}
  {2001})}\BibitemShut {NoStop}%
\bibitem [{\citenamefont {Moline}\ \emph {et~al.}(2017)\citenamefont {Moline},
  \citenamefont {Sanchez-Conde}, \citenamefont {Palomares-Ruiz},\ and\
  \citenamefont {Prada}}]{moline2017characterization}%
  \BibitemOpen
  \bibfield  {author} {\bibinfo {author} {\bibfnamefont {A.}~\bibnamefont
  {Moline}}, \bibinfo {author} {\bibfnamefont {M.}~\bibnamefont
  {Sanchez-Conde}}, \bibinfo {author} {\bibfnamefont {S.}~\bibnamefont
  {Palomares-Ruiz}}, \ and\ \bibinfo {author} {\bibfnamefont {F.}~\bibnamefont
  {Prada}},\ }\href@noop {} {\bibfield  {journal} {\bibinfo  {journal} {Monthly
  Notices of the Royal Astronomical Society}\ }\textbf {\bibinfo {volume}
  {466}},\ \bibinfo {pages} {4974} (\bibinfo {year} {2017})}\BibitemShut
  {NoStop}%
\bibitem [{\citenamefont {Navarro}\ \emph {et~al.}(1996)\citenamefont
  {Navarro}, \citenamefont {Frenk},\ and\ \citenamefont
  {White}}]{Navarro:1995iw}%
  \BibitemOpen
  \bibfield  {author} {\bibinfo {author} {\bibfnamefont {J.~F.}\ \bibnamefont
  {Navarro}}, \bibinfo {author} {\bibfnamefont {C.~S.}\ \bibnamefont {Frenk}},
  \ and\ \bibinfo {author} {\bibfnamefont {S.~D.~M.}\ \bibnamefont {White}},\
  }\href {\doibase 10.1086/177173} {\bibfield  {journal} {\bibinfo  {journal}
  {Astrophys. J.}\ }\textbf {\bibinfo {volume} {462}},\ \bibinfo {pages} {563}
  (\bibinfo {year} {1996})},\ \Eprint {http://arxiv.org/abs/astro-ph/9508025}
  {arXiv:astro-ph/9508025 [astro-ph]} \BibitemShut {NoStop}%
%%CITATION = ASTRO-PH/9508025;%%
\bibitem [{\citenamefont {Burkert}(1996)}]{Burkert:1995yz}%
  \BibitemOpen
  \bibfield  {author} {\bibinfo {author} {\bibfnamefont {A.}~\bibnamefont
  {Burkert}},\ }\bibfield  {booktitle} {\emph {\bibinfo {booktitle} {{IAU
  Symposium 171: New Light on Galaxy Evolution Heidelberg, Germany, June 26-30,
  1995}}},\ }\href {\doibase 10.1086/309560} {\bibfield  {journal} {\bibinfo
  {journal} {IAU Symp.}\ }\textbf {\bibinfo {volume} {171}},\ \bibinfo {pages}
  {175} (\bibinfo {year} {1996})},\ \bibinfo {note} {[Astrophys.
  J.447,L25(1995)]},\ \Eprint {http://arxiv.org/abs/astro-ph/9504041}
  {arXiv:astro-ph/9504041 [astro-ph]} \BibitemShut {NoStop}%
%%CITATION = ASTRO-PH/9504041;%%
\bibitem [{\citenamefont {Strigari}\ \emph {et~al.}(2007)\citenamefont
  {Strigari} \emph {et~al.}}]{strigari2007redefining}%
  \BibitemOpen
  \bibfield  {author} {\bibinfo {author} {\bibfnamefont {L.~E.}\ \bibnamefont
  {Strigari}} \emph {et~al.},\ }\href@noop {} {\bibfield  {journal} {\bibinfo
  {journal} {The Astrophysical Journal}\ }\textbf {\bibinfo {volume} {669}},\
  \bibinfo {pages} {676} (\bibinfo {year} {2007})}\BibitemShut {NoStop}%
\bibitem [{\citenamefont {Walker}\ \emph {et~al.}(2007)\citenamefont {Walker}
  \emph {et~al.}}]{walker2007velocity}%
  \BibitemOpen
  \bibfield  {author} {\bibinfo {author} {\bibfnamefont {M.~G.}\ \bibnamefont
  {Walker}} \emph {et~al.},\ }\href@noop {} {\bibfield  {journal} {\bibinfo
  {journal} {The Astrophysical Journal Letters}\ }\textbf {\bibinfo {volume}
  {667}},\ \bibinfo {pages} {L53} (\bibinfo {year} {2007})}\BibitemShut
  {NoStop}%
\bibitem [{\citenamefont {Navarro}\ \emph {et~al.}(2010)\citenamefont
  {Navarro}, \citenamefont {Ludlow}, \citenamefont {Springel}, \citenamefont
  {Wang}, \citenamefont {Vogelsberger}, \citenamefont {White}, \citenamefont
  {Jenkins}, \citenamefont {Frenk},\ and\ \citenamefont
  {Helmi}}]{Navarro:2008kc}%
  \BibitemOpen
  \bibfield  {author} {\bibinfo {author} {\bibfnamefont {J.~F.}\ \bibnamefont
  {Navarro}}, \bibinfo {author} {\bibfnamefont {A.}~\bibnamefont {Ludlow}},
  \bibinfo {author} {\bibfnamefont {V.}~\bibnamefont {Springel}}, \bibinfo
  {author} {\bibfnamefont {J.}~\bibnamefont {Wang}}, \bibinfo {author}
  {\bibfnamefont {M.}~\bibnamefont {Vogelsberger}}, \bibinfo {author}
  {\bibfnamefont {S.~D.~M.}\ \bibnamefont {White}}, \bibinfo {author}
  {\bibfnamefont {A.}~\bibnamefont {Jenkins}}, \bibinfo {author} {\bibfnamefont
  {C.~S.}\ \bibnamefont {Frenk}}, \ and\ \bibinfo {author} {\bibfnamefont
  {A.}~\bibnamefont {Helmi}},\ }\href {\doibase
  10.1111/j.1365-2966.2009.15878.x} {\bibfield  {journal} {\bibinfo  {journal}
  {Mon. Not. Roy. Astron. Soc.}\ }\textbf {\bibinfo {volume} {402}},\ \bibinfo
  {pages} {21} (\bibinfo {year} {2010})},\ \Eprint
  {http://arxiv.org/abs/0810.1522} {arXiv:0810.1522 [astro-ph]} \BibitemShut
  {NoStop}%
%%CITATION = ARXIV:0810.1522;%%
\bibitem [{\citenamefont {Simon}\ \emph {et~al.}(2005)\citenamefont {Simon},
  \citenamefont {Bolatto}, \citenamefont {Leroy}, \citenamefont {Blitz},\ and\
  \citenamefont {Gates}}]{Simon:2004sr}%
  \BibitemOpen
  \bibfield  {author} {\bibinfo {author} {\bibfnamefont {J.~D.}\ \bibnamefont
  {Simon}}, \bibinfo {author} {\bibfnamefont {A.~D.}\ \bibnamefont {Bolatto}},
  \bibinfo {author} {\bibfnamefont {A.}~\bibnamefont {Leroy}}, \bibinfo
  {author} {\bibfnamefont {L.}~\bibnamefont {Blitz}}, \ and\ \bibinfo {author}
  {\bibfnamefont {E.~L.}\ \bibnamefont {Gates}},\ }\href {\doibase
  10.1086/427684} {\bibfield  {journal} {\bibinfo  {journal} {Astrophys. J.}\
  }\textbf {\bibinfo {volume} {621}},\ \bibinfo {pages} {757} (\bibinfo {year}
  {2005})},\ \Eprint {http://arxiv.org/abs/astro-ph/0412035}
  {arXiv:astro-ph/0412035 [astro-ph]} \BibitemShut {NoStop}%
%%CITATION = ASTRO-PH/0412035;%%
\bibitem [{\citenamefont {Weldrake}\ \emph {et~al.}(2003)\citenamefont
  {Weldrake}, \citenamefont {de~Blok},\ and\ \citenamefont
  {Walter}}]{Weldrake:2002ri}%
  \BibitemOpen
  \bibfield  {author} {\bibinfo {author} {\bibfnamefont {D.}~\bibnamefont
  {Weldrake}}, \bibinfo {author} {\bibfnamefont {E.}~\bibnamefont {de~Blok}}, \
  and\ \bibinfo {author} {\bibfnamefont {F.}~\bibnamefont {Walter}},\ }\href
  {\doibase 10.1046/j.1365-8711.2003.06170.x} {\bibfield  {journal} {\bibinfo
  {journal} {Mon. Not. Roy. Astron. Soc.}\ }\textbf {\bibinfo {volume} {340}},\
  \bibinfo {pages} {12} (\bibinfo {year} {2003})},\ \Eprint
  {http://arxiv.org/abs/astro-ph/0210568} {arXiv:astro-ph/0210568 [astro-ph]}
  \BibitemShut {NoStop}%
%%CITATION = ASTRO-PH/0210568;%%
\bibitem [{\citenamefont {{Chemin}}\ \emph {et~al.}(2011)\citenamefont
  {{Chemin}}, \citenamefont {{de Blok}},\ and\ \citenamefont
  {{Mamon}}}]{2011AJ....142..109C}%
  \BibitemOpen
  \bibfield  {author} {\bibinfo {author} {\bibfnamefont {L.}~\bibnamefont
  {{Chemin}}}, \bibinfo {author} {\bibfnamefont {W.~J.~G.}\ \bibnamefont {{de
  Blok}}}, \ and\ \bibinfo {author} {\bibfnamefont {G.~A.}\ \bibnamefont
  {{Mamon}}},\ }\href {\doibase 10.1088/0004-6256/142/4/109} {\bibfield
  {journal} {\bibinfo  {journal} {\aj}\ }\textbf {\bibinfo {volume} {142}},\
  \bibinfo {eid} {109} (\bibinfo {year} {2011})},\ \Eprint
  {http://arxiv.org/abs/1109.4247} {arXiv:1109.4247 [astro-ph.CO]} \BibitemShut
  {NoStop}%
\bibitem [{\citenamefont {Gnedin}\ \emph {et~al.}(2011)\citenamefont {Gnedin}
  \emph {et~al.}}]{gnedin2011halo}%
  \BibitemOpen
  \bibfield  {author} {\bibinfo {author} {\bibfnamefont {O.}~\bibnamefont
  {Gnedin}} \emph {et~al.},\ }\href@noop {} {\bibfield  {journal} {\bibinfo
  {journal} {arXiv preprint arXiv:1108.5736}\ } (\bibinfo {year}
  {2011})}\BibitemShut {NoStop}%
\bibitem [{\citenamefont {Seigar}\ \emph {et~al.}(2008)\citenamefont {Seigar},
  \citenamefont {Barth},\ and\ \citenamefont {Bullock}}]{seigar2008revised}%
  \BibitemOpen
  \bibfield  {author} {\bibinfo {author} {\bibfnamefont {M.~S.}\ \bibnamefont
  {Seigar}}, \bibinfo {author} {\bibfnamefont {A.~J.}\ \bibnamefont {Barth}}, \
  and\ \bibinfo {author} {\bibfnamefont {J.~S.}\ \bibnamefont {Bullock}},\
  }\href@noop {} {\bibfield  {journal} {\bibinfo  {journal} {Monthly Notices of
  the Royal Astronomical Society}\ }\textbf {\bibinfo {volume} {389}},\
  \bibinfo {pages} {1911} (\bibinfo {year} {2008})}\BibitemShut {NoStop}%
\bibitem [{\citenamefont {Rubin}\ and\ \citenamefont
  {Ford}(1970)}]{Rubin:1970zza}%
  \BibitemOpen
  \bibfield  {author} {\bibinfo {author} {\bibfnamefont {V.~C.}\ \bibnamefont
  {Rubin}}\ and\ \bibinfo {author} {\bibfnamefont {W.~K.}\ \bibnamefont {Ford},
  \bibfnamefont {Jr.}},\ }\href {\doibase 10.1086/150317} {\bibfield  {journal}
  {\bibinfo  {journal} {Astrophys. J.}\ }\textbf {\bibinfo {volume} {159}},\
  \bibinfo {pages} {379} (\bibinfo {year} {1970})}\BibitemShut {NoStop}%
%%CITATION = ASJOA,159,379;%%
\bibitem [{\citenamefont {Binder}\ \emph {et~al.}(2017)\citenamefont {Binder},
  \citenamefont {Bringmann}, \citenamefont {Gustafsson},\ and\ \citenamefont
  {Hryczuk}}]{binder2017early}%
  \BibitemOpen
  \bibfield  {author} {\bibinfo {author} {\bibfnamefont {T.}~\bibnamefont
  {Binder}}, \bibinfo {author} {\bibfnamefont {T.}~\bibnamefont {Bringmann}},
  \bibinfo {author} {\bibfnamefont {M.}~\bibnamefont {Gustafsson}}, \ and\
  \bibinfo {author} {\bibfnamefont {A.}~\bibnamefont {Hryczuk}},\ }\href@noop
  {} {\bibfield  {journal} {\bibinfo  {journal} {arXiv preprint
  arXiv:1706.07433}\ } (\bibinfo {year} {2017})}\BibitemShut {NoStop}%
\bibitem [{\citenamefont {Bringmann}(2009)}]{bringmann2009particle}%
  \BibitemOpen
  \bibfield  {author} {\bibinfo {author} {\bibfnamefont {T.}~\bibnamefont
  {Bringmann}},\ }\href@noop {} {\bibfield  {journal} {\bibinfo  {journal} {New
  Journal of Physics}\ }\textbf {\bibinfo {volume} {11}},\ \bibinfo {pages}
  {105027} (\bibinfo {year} {2009})}\BibitemShut {NoStop}%
\bibitem [{\citenamefont {Abeysekara}\ \emph {et~al.}(2017)\citenamefont
  {Abeysekara} \emph {et~al.}}]{Abeysekara:2017mjj}%
  \BibitemOpen
  \bibfield  {author} {\bibinfo {author} {\bibfnamefont {A.~U.}\ \bibnamefont
  {Abeysekara}} \emph {et~al.},\ }\href {\doibase 10.3847/1538-4357/aa7555}
  {\bibfield  {journal} {\bibinfo  {journal} {Astrophys. J.}\ }\textbf
  {\bibinfo {volume} {843}},\ \bibinfo {pages} {39} (\bibinfo {year} {2017})},\
  \Eprint {http://arxiv.org/abs/1701.01778} {arXiv:1701.01778 [astro-ph.HE]}
  \BibitemShut {NoStop}%
%%CITATION = ARXIV:1701.01778;%%
\bibitem [{\citenamefont {{Abdo}}\ \emph {et~al.}(2012)\citenamefont {{Abdo}}
  \emph {et~al.}}]{2012ApJ...750...63A}%
  \BibitemOpen
  \bibfield  {author} {\bibinfo {author} {\bibfnamefont {A.~A.}\ \bibnamefont
  {{Abdo}}} \emph {et~al.} (\bibinfo {collaboration} {Milagro}),\ }\href
  {\doibase 10.1088/0004-637X/750/1/63} {\bibfield  {journal} {\bibinfo
  {journal} {\apj}\ }\textbf {\bibinfo {volume} {750}},\ \bibinfo {eid} {63}
  (\bibinfo {year} {2012})},\ \Eprint {http://arxiv.org/abs/1110.0409}
  {arXiv:1110.0409 [astro-ph.HE]} \BibitemShut {NoStop}%
\bibitem [{\citenamefont {Olive}\ \emph {et~al.}(2014)\citenamefont {Olive}
  \emph {et~al.}}]{Agashe:2014kda}%
  \BibitemOpen
  \bibfield  {author} {\bibinfo {author} {\bibfnamefont {K.~A.}\ \bibnamefont
  {Olive}} \emph {et~al.} (\bibinfo {collaboration} {Particle Data Group}),\
  }\href {\doibase 10.1088/1674-1137/38/9/090001} {\bibfield  {journal}
  {\bibinfo  {journal} {Chin. Phys.}\ }\textbf {\bibinfo {volume} {C38}},\
  \bibinfo {pages} {090001} (\bibinfo {year} {2014})}\BibitemShut {NoStop}%
%%CITATION = CHPHD,C38,090001;%%
\bibitem [{\citenamefont {Rolke}\ \emph {et~al.}(2005)\citenamefont {Rolke},
  \citenamefont {Lopez},\ and\ \citenamefont {Conrad}}]{Rolke:2004mj}%
  \BibitemOpen
  \bibfield  {author} {\bibinfo {author} {\bibfnamefont {W.~A.}\ \bibnamefont
  {Rolke}}, \bibinfo {author} {\bibfnamefont {A.~M.}\ \bibnamefont {Lopez}}, \
  and\ \bibinfo {author} {\bibfnamefont {J.}~\bibnamefont {Conrad}},\ }\href
  {\doibase 10.1016/j.nima.2005.05.068} {\bibfield  {journal} {\bibinfo
  {journal} {Nucl. Instrum. Meth.}\ }\textbf {\bibinfo {volume} {A551}},\
  \bibinfo {pages} {493} (\bibinfo {year} {2005})},\ \Eprint
  {http://arxiv.org/abs/physics/0403059} {arXiv:physics/0403059 [physics]}
  \BibitemShut {NoStop}%
%%CITATION = PHYSICS/0403059;%%
\bibitem [{\citenamefont {Venters}(2010)}]{Venters:2010bq}%
  \BibitemOpen
  \bibfield  {author} {\bibinfo {author} {\bibfnamefont {T.~M.}\ \bibnamefont
  {Venters}},\ }\href {\doibase 10.1088/0004-637X/710/2/1530} {\bibfield
  {journal} {\bibinfo  {journal} {Astrophys. J.}\ }\textbf {\bibinfo {volume}
  {710}},\ \bibinfo {pages} {1530} (\bibinfo {year} {2010})},\ \Eprint
  {http://arxiv.org/abs/1001.1363} {arXiv:1001.1363 [astro-ph.HE]} \BibitemShut
  {NoStop}%
%%CITATION = ARXIV:1001.1363;%%
\bibitem [{\citenamefont {Moskalenko}\ \emph {et~al.}(2006)\citenamefont
  {Moskalenko}, \citenamefont {Porter},\ and\ \citenamefont
  {Strong}}]{Moskalenko:2005ng}%
  \BibitemOpen
  \bibfield  {author} {\bibinfo {author} {\bibfnamefont {I.~V.}\ \bibnamefont
  {Moskalenko}}, \bibinfo {author} {\bibfnamefont {T.~A.}\ \bibnamefont
  {Porter}}, \ and\ \bibinfo {author} {\bibfnamefont {A.~W.}\ \bibnamefont
  {Strong}},\ }\href {\doibase 10.1086/503524} {\bibfield  {journal} {\bibinfo
  {journal} {Astrophys. J.}\ }\textbf {\bibinfo {volume} {640}},\ \bibinfo
  {pages} {L155} (\bibinfo {year} {2006})},\ \Eprint
  {http://arxiv.org/abs/astro-ph/0511149} {arXiv:astro-ph/0511149 [astro-ph]}
  \BibitemShut {NoStop}%
%%CITATION = ASTRO-PH/0511149;%%
\bibitem [{\citenamefont {Marinelli}\ and\ \citenamefont
  {Goodman}(2017)}]{Marinelli:2017vzu}%
  \BibitemOpen
  \bibfield  {author} {\bibinfo {author} {\bibfnamefont {S.~S.}\ \bibnamefont
  {Marinelli}}\ and\ \bibinfo {author} {\bibfnamefont {J.}~\bibnamefont
  {Goodman}} (\bibinfo {collaboration} {HAWC}),\ }in\ \href
  {https://inspirehep.net/record/1615737/files/arXiv:1708.03502.pdf} {\emph
  {\bibinfo {booktitle} {{Proceedings, 35th International Cosmic Ray Conference
  (ICRC 2017): Bexco, Busan, Korea, July 12-20, 2017}}}}\ (\bibinfo {year}
  {2017})\ \Eprint {http://arxiv.org/abs/1708.03502} {arXiv:1708.03502
  [astro-ph.IM]} \BibitemShut {NoStop}%
%%CITATION = ARXIV:1708.03502;%%
\bibitem [{\citenamefont {Aartsen}\ \emph {et~al.}(2017)\citenamefont {Aartsen}
  \emph {et~al.}}]{Aartsen:2017snx}%
  \BibitemOpen
  \bibfield  {author} {\bibinfo {author} {\bibfnamefont {M.~G.}\ \bibnamefont
  {Aartsen}} \emph {et~al.},\ }\href@noop {} {\  (\bibinfo {year} {2017})},\
  \Eprint {http://arxiv.org/abs/1702.06131} {arXiv:1702.06131 [astro-ph.HE]}
  \BibitemShut {NoStop}%
%%CITATION = ARXIV:1702.06131;%%
\bibitem [{\citenamefont {Bird}(2016)}]{Bird:2015npa}%
  \BibitemOpen
  \bibfield  {author} {\bibinfo {author} {\bibfnamefont {R.}~\bibnamefont
  {Bird}} (\bibinfo {collaboration} {VERITAS}),\ }\bibfield  {booktitle} {\emph
  {\bibinfo {booktitle} {{Proceedings, 34th International Cosmic Ray Conference
  (ICRC 2015): The Hague, The Netherlands, July 30-August 6, 2015}}},\
  }\href@noop {} {\bibfield  {journal} {\bibinfo  {journal} {PoS}\ }\textbf
  {\bibinfo {volume} {ICRC2015}},\ \bibinfo {pages} {851} (\bibinfo {year}
  {2016})},\ \Eprint {http://arxiv.org/abs/1508.07195} {arXiv:1508.07195
  [astro-ph.HE]} \BibitemShut {NoStop}%
%%CITATION = ARXIV:1508.07195;%%
\bibitem [{\citenamefont {Harding}\ \emph {et~al.}(shed)\citenamefont {Harding}
  \emph {et~al.}}]{harding}%
  \BibitemOpen
  \bibfield  {author} {\bibinfo {author} {\bibfnamefont {J.~P.}\ \bibnamefont
  {Harding}} \emph {et~al.},\ }\href@noop {} {\bibfield  {journal} {\bibinfo
  {journal} {Astrophys. J.}\ } (\bibinfo {year} {\noop{2017}to be
  published})}\BibitemShut {NoStop}%
\bibitem [{\citenamefont {Abdallah}\ \emph {et~al.}(2016)\citenamefont
  {Abdallah} \emph {et~al.}}]{Abdallah:2016ygi}%
  \BibitemOpen
  \bibfield  {author} {\bibinfo {author} {\bibfnamefont {H.}~\bibnamefont
  {Abdallah}} \emph {et~al.} (\bibinfo {collaboration} {H.E.S.S.}),\ }\href
  {\doibase 10.1103/PhysRevLett.117.111301} {\bibfield  {journal} {\bibinfo
  {journal} {Phys. Rev. Lett.}\ }\textbf {\bibinfo {volume} {117}},\ \bibinfo
  {pages} {111301} (\bibinfo {year} {2016})},\ \Eprint
  {http://arxiv.org/abs/1607.08142} {arXiv:1607.08142 [astro-ph.HE]}
  \BibitemShut {NoStop}%
%%CITATION = ARXIV:1607.08142;%%
\bibitem [{\citenamefont {Aliu}\ \emph {et~al.}(2012)\citenamefont {Aliu} \emph
  {et~al.}}]{Aliu:2012ga}%
  \BibitemOpen
  \bibfield  {author} {\bibinfo {author} {\bibfnamefont {E.}~\bibnamefont
  {Aliu}} \emph {et~al.} (\bibinfo {collaboration} {VERITAS}),\ }\href
  {\doibase 10.1103/PhysRevD.85.062001, 10.1103/PhysRevD.91.129903} {\bibfield
  {journal} {\bibinfo  {journal} {Phys. Rev.}\ }\textbf {\bibinfo {volume}
  {D85}},\ \bibinfo {pages} {062001} (\bibinfo {year} {2012})},\ \bibinfo
  {note} {[Erratum: Phys. Rev.D91,no.12,129903(2015)]},\ \Eprint
  {http://arxiv.org/abs/1202.2144} {arXiv:1202.2144 [astro-ph.HE]} \BibitemShut
  {NoStop}%
%%CITATION = ARXIV:1202.2144;%%
\bibitem [{\citenamefont {Baring}\ \emph {et~al.}(2016)\citenamefont {Baring},
  \citenamefont {Ghosh}, \citenamefont {Queiroz},\ and\ \citenamefont
  {Sinha}}]{Baring:2015sza}%
  \BibitemOpen
  \bibfield  {author} {\bibinfo {author} {\bibfnamefont {M.~G.}\ \bibnamefont
  {Baring}}, \bibinfo {author} {\bibfnamefont {T.}~\bibnamefont {Ghosh}},
  \bibinfo {author} {\bibfnamefont {F.~S.}\ \bibnamefont {Queiroz}}, \ and\
  \bibinfo {author} {\bibfnamefont {K.}~\bibnamefont {Sinha}},\ }\href
  {\doibase 10.1103/PhysRevD.93.103009} {\bibfield  {journal} {\bibinfo
  {journal} {Phys. Rev.}\ }\textbf {\bibinfo {volume} {D93}},\ \bibinfo {pages}
  {103009} (\bibinfo {year} {2016})},\ \Eprint
  {http://arxiv.org/abs/1510.00389} {arXiv:1510.00389 [hep-ph]} \BibitemShut
  {NoStop}%
%%CITATION = ARXIV:1510.00389;%%
\bibitem [{\citenamefont {Esmaili}\ \emph {et~al.}(2014)\citenamefont
  {Esmaili}, \citenamefont {Kang},\ and\ \citenamefont
  {Serpico}}]{Esmaili:2014rma}%
  \BibitemOpen
  \bibfield  {author} {\bibinfo {author} {\bibfnamefont {A.}~\bibnamefont
  {Esmaili}}, \bibinfo {author} {\bibfnamefont {S.~K.}\ \bibnamefont {Kang}}, \
  and\ \bibinfo {author} {\bibfnamefont {P.~D.}\ \bibnamefont {Serpico}},\
  }\href {\doibase 10.1088/1475-7516/2014/12/054} {\bibfield  {journal}
  {\bibinfo  {journal} {JCAP}\ }\textbf {\bibinfo {volume} {1412}},\ \bibinfo
  {pages} {054} (\bibinfo {year} {2014})},\ \Eprint
  {http://arxiv.org/abs/1410.5979} {arXiv:1410.5979 [hep-ph]} \BibitemShut
  {NoStop}%
%%CITATION = ARXIV:1410.5979;%%
\bibitem [{\citenamefont {Archambault}\ \emph {et~al.}(2017)\citenamefont
  {Archambault} \emph {et~al.}}]{Archambault:2017wyh}%
  \BibitemOpen
  \bibfield  {author} {\bibinfo {author} {\bibfnamefont {S.}~\bibnamefont
  {Archambault}} \emph {et~al.} (\bibinfo {collaboration} {VERITAS}),\ }\href
  {\doibase 10.1103/PhysRevD.95.082001} {\bibfield  {journal} {\bibinfo
  {journal} {Phys. Rev.}\ }\textbf {\bibinfo {volume} {D95}},\ \bibinfo {pages}
  {082001} (\bibinfo {year} {2017})},\ \Eprint
  {http://arxiv.org/abs/1703.04937} {arXiv:1703.04937 [astro-ph.HE]}
  \BibitemShut {NoStop}%
%%CITATION = ARXIV:1703.04937;%%
\bibitem [{\citenamefont {Aleksić}\ \emph {et~al.}(2014)\citenamefont
  {Aleksić} \emph {et~al.}}]{Aleksic:2013xea}%
  \BibitemOpen
  \bibfield  {author} {\bibinfo {author} {\bibfnamefont {J.}~\bibnamefont
  {Aleksić}} \emph {et~al.},\ }\href {\doibase 10.1088/1475-7516/2014/02/008}
  {\bibfield  {journal} {\bibinfo  {journal} {JCAP}\ }\textbf {\bibinfo
  {volume} {1402}},\ \bibinfo {pages} {008} (\bibinfo {year} {2014})},\ \Eprint
  {http://arxiv.org/abs/1312.1535} {arXiv:1312.1535 [hep-ph]} \BibitemShut
  {NoStop}%
%%CITATION = ARXIV:1312.1535;%%
\bibitem [{\citenamefont {Aartsen}\ \emph {et~al.}(2013)\citenamefont {Aartsen}
  \emph {et~al.}}]{Aartsen:2013dxa}%
  \BibitemOpen
  \bibfield  {author} {\bibinfo {author} {\bibfnamefont {M.~G.}\ \bibnamefont
  {Aartsen}} \emph {et~al.} (\bibinfo {collaboration} {IceCube}),\ }\href
  {\doibase 10.1103/PhysRevD.88.122001} {\bibfield  {journal} {\bibinfo
  {journal} {Phys. Rev.}\ }\textbf {\bibinfo {volume} {D88}},\ \bibinfo {pages}
  {122001} (\bibinfo {year} {2013})},\ \Eprint {http://arxiv.org/abs/1307.3473}
  {arXiv:1307.3473 [astro-ph.HE]} \BibitemShut {NoStop}%
%%CITATION = ARXIV:1307.3473;%%
\bibitem [{\citenamefont {Aartsen}\ \emph {et~al.}(2018)\citenamefont {Aartsen}
  \emph {et~al.}}]{Aartsen:2018mxl}%
  \BibitemOpen
  \bibfield  {author} {\bibinfo {author} {\bibfnamefont {M.~G.}\ \bibnamefont
  {Aartsen}} \emph {et~al.} (\bibinfo {collaboration} {IceCube}),\ }\href@noop
  {} {\  (\bibinfo {year} {2018})},\ \Eprint {http://arxiv.org/abs/1804.03848}
  {arXiv:1804.03848 [astro-ph.HE]} \BibitemShut {NoStop}%
%%CITATION = ARXIV:1804.03848;%%
\end{thebibliography}%
